\documentclass[10pt,twoside,twocolumn]{IEEEtran}

\usepackage{cite}
\usepackage[T1]{fontenc}
\usepackage{graphicx}
\usepackage{amssymb}
\usepackage{amsmath}
\usepackage{amsthm}
\usepackage{subfigure}
\usepackage{booktabs} 
\usepackage{multirow}
\usepackage{microtype}
\usepackage{balance}
\usepackage{xcolor}
\usepackage{xfrac}    
\usepackage{algorithm}
\usepackage{setspace}

\usepackage{algorithmicx}
\usepackage{algpseudocode}
\algrenewcommand\algorithmicindent{0.7em}%
\usepackage{xpatch}

\usepackage{tikz}

\usepackage{hyperref}

\usepackage{amssymb}
\usepackage{amsfonts}
\usepackage{mathrsfs}
\usepackage{xspace}
\usepackage{bm}
\usepackage{upgreek}

\newcommand{\safemath}[2]{\newcommand{#1}{\ensuremath{#2}\xspace}}

\safemath{\bma}{\mathbf{a}}
\safemath{\bmb}{\mathbf{b}}
\safemath{\bmc}{\mathbf{c}}
\safemath{\bmd}{\mathbf{d}}
\safemath{\bme}{\mathbf{e}}
\safemath{\bmf}{\mathbf{f}}
\safemath{\bmg}{\mathbf{g}}
\safemath{\bmh}{\mathbf{h}}
\safemath{\bmi}{\mathbf{i}}
\safemath{\bmj}{\mathbf{j}}
\safemath{\bmk}{\mathbf{k}}
\safemath{\bml}{\mathbf{l}}
\safemath{\bmm}{\mathbf{m}}
\safemath{\bmn}{\mathbf{n}}
\safemath{\bmo}{\mathbf{o}}
\safemath{\bmp}{\mathbf{p}}
\safemath{\bmq}{\mathbf{q}}
\safemath{\bmr}{\mathbf{r}}
\safemath{\bms}{\mathbf{s}}
\safemath{\bmt}{\mathbf{t}}
\safemath{\bmu}{\mathbf{u}}
\safemath{\bmv}{\mathbf{v}}
\safemath{\bmw}{\mathbf{w}}
\safemath{\bmx}{\mathbf{x}}
\safemath{\bmy}{\mathbf{y}}
\safemath{\bmz}{\mathbf{z}}
\safemath{\bmzero}{\mathbf{0}}
\safemath{\bmone}{\mathbf{1}}

\bmdefine{\biad}{a}
\bmdefine{\bibd}{b}
\bmdefine{\bicd}{c}
\bmdefine{\bidd}{d}
\bmdefine{\bied}{e}
\bmdefine{\bifd}{f}
\bmdefine{\bigd}{g}
\bmdefine{\bihd}{h}
\bmdefine{\biid}{i}
\bmdefine{\bijd}{j}
\bmdefine{\bikd}{k}
\bmdefine{\bild}{l}
\bmdefine{\bimd}{m}
\bmdefine{\bind}{n}
\bmdefine{\biod}{o}
\bmdefine{\bipd}{p}
\bmdefine{\biqd}{q}
\bmdefine{\bird}{r}
\bmdefine{\bisd}{s}
\bmdefine{\bitd}{t}
\bmdefine{\biud}{u}
\bmdefine{\bivd}{v}
\bmdefine{\biwd}{w}
\bmdefine{\bixd}{x}
\bmdefine{\biyd}{y}
\bmdefine{\bizd}{z}

\bmdefine{\bixid}{\xi}
\bmdefine{\bilambdad}{\lambda}
\bmdefine{\bimud}{\mu}
\bmdefine{\bithetad}{\theta}
\bmdefine{\biphid}{\phi}
\bmdefine{\bideltad}{\delta}

\safemath{\bmia}{\biad}
\safemath{\bmib}{\bibd}
\safemath{\bmic}{\bicd}
\safemath{\bmid}{\bidd}
\safemath{\bmie}{\bied}
\safemath{\bmif}{\bifd}
\safemath{\bmig}{\bigd}
\safemath{\bmih}{\bihd}
\safemath{\bmii}{\biid}
\safemath{\bmij}{\bijd}
\safemath{\bmik}{\bikd}
\safemath{\bmil}{\bild}
\safemath{\bmim}{\bimd}
\safemath{\bmin}{\bind}
\safemath{\bmio}{\biod}
\safemath{\bmip}{\bipd}
\safemath{\bmiq}{\biqd}
\safemath{\bmir}{\bird}
\safemath{\bmis}{\bisd}
\safemath{\bmit}{\bitd}
\safemath{\bmiu}{\biud}
\safemath{\bmiv}{\bivd}
\safemath{\bmiw}{\biwd}
\safemath{\bmix}{\bixd}
\safemath{\bmiy}{\biyd}
\safemath{\bmiz}{\bizd}

\safemath{\bmxi}{\bixid}
\safemath{\bmlambda}{\bilambdad}
\safemath{\bmmu}{\bimud}
\safemath{\bmtheta}{\bithetad}
\safemath{\bmphi}{\biphid}
\safemath{\bmdelta}{\bideltad}

\safemath{\bA}{\mathbf{A}}
\safemath{\bB}{\mathbf{B}}
\safemath{\bC}{\mathbf{C}}
\safemath{\bD}{\mathbf{D}}
\safemath{\bE}{\mathbf{E}}
\safemath{\bF}{\mathbf{F}}
\safemath{\bG}{\mathbf{G}}
\safemath{\bH}{\mathbf{H}}
\safemath{\bI}{\mathbf{I}}
\safemath{\bJ}{\mathbf{J}}
\safemath{\bK}{\mathbf{K}}
\safemath{\bL}{\mathbf{L}}
\safemath{\bM}{\mathbf{M}}
\safemath{\bN}{\mathbf{N}}
\safemath{\bO}{\mathbf{O}}
\safemath{\bP}{\mathbf{P}}
\safemath{\bQ}{\mathbf{Q}}
\safemath{\bR}{\mathbf{R}}
\safemath{\bS}{\mathbf{S}}
\safemath{\bT}{\mathbf{T}}
\safemath{\bU}{\mathbf{U}}
\safemath{\bV}{\mathbf{V}}
\safemath{\bW}{\mathbf{W}}
\safemath{\bX}{\mathbf{X}}
\safemath{\bY}{\mathbf{Y}}
\safemath{\bZ}{\mathbf{Z}}

\safemath{\bZero}{\mathbf{0}}
\safemath{\bOne}{\mathbf{1}}
\safemath{\bDelta}{\mathbf{\Delta}}
\safemath{\bLambda}{\mathbf{\UpLambda}}
\safemath{\bPhi}{\mathbf{\Upphi}}
\safemath{\bSigma}{\mathbf{\Upsigma}}
\safemath{\bOmega}{\mathbf{\Upomega}}
\safemath{\bTheta}{\mathbf{\Uptheta}}

\bmdefine{\biAd}{A}
\bmdefine{\biBd}{B}
\bmdefine{\biCd}{C}
\bmdefine{\biDd}{D}
\bmdefine{\biEd}{E}
\bmdefine{\biFd}{F}
\bmdefine{\biGd}{G}
\bmdefine{\biHd}{H}
\bmdefine{\biId}{I}
\bmdefine{\biJd}{J}
\bmdefine{\biKd}{K}
\bmdefine{\biLd}{L}
\bmdefine{\biMd}{M}
\bmdefine{\biOd}{N}
\bmdefine{\biPd}{O}
\bmdefine{\biQd}{P}
\bmdefine{\biRd}{R}
\bmdefine{\biSd}{S}
\bmdefine{\biTd}{T}
\bmdefine{\biUd}{U}
\bmdefine{\biVd}{V}
\bmdefine{\biWd}{W}
\bmdefine{\biXd}{X}
\bmdefine{\biYd}{Y}
\bmdefine{\biZd}{Z}

\bmdefine{\biDelta}{\Delta}
\bmdefine{\biLambda}{\Lambda}
\bmdefine{\biPhi}{\Phi}
\bmdefine{\biSigma}{\Sigma}
\bmdefine{\biOmega}{\Omega}
\bmdefine{\biTheta}{\Theta}

\safemath{\bimA}{\biAd}
\safemath{\bimB}{\biBd}
\safemath{\bimC}{\biCd}
\safemath{\bimD}{\biDd}
\safemath{\bimE}{\biEd}
\safemath{\bimF}{\biFd}
\safemath{\bimG}{\biGd}
\safemath{\bimH}{\biHd}
\safemath{\bimI}{\biId}
\safemath{\bimJ}{\biJd}
\safemath{\bimK}{\biKd}
\safemath{\bimL}{\biLd}
\safemath{\bimM}{\biMd}
\safemath{\bimN}{\biNd}
\safemath{\bimO}{\biOd}
\safemath{\bimP}{\biPd}
\safemath{\bimQ}{\biQd}
\safemath{\bimR}{\biRd}
\safemath{\bimS}{\biSd}
\safemath{\bimT}{\biTd}
\safemath{\bimU}{\biUd}
\safemath{\bimV}{\biVd}
\safemath{\bimW}{\biWd}
\safemath{\bimX}{\biXd}
\safemath{\bimY}{\biYd}
\safemath{\bimZ}{\biZd}

\safemath{\bimDelta}{\biDelta}
\safemath{\bimLambda}{\biLambda}
\safemath{\bimPhi}{\biPhi}
\safemath{\bimSigma}{\biSigma}
\safemath{\bimOmega}{\biOmega}
\safemath{\bimTheta}{\biTheta}

\safemath{\setA}{\mathcal{A}}
\safemath{\setB}{\mathcal{B}}
\safemath{\setC}{\mathcal{C}}
\safemath{\setD}{\mathcal{D}}
\safemath{\setE}{\mathcal{E}}
\safemath{\setF}{\mathcal{F}}
\safemath{\setG}{\mathcal{G}}
\safemath{\setH}{\mathcal{H}}
\safemath{\setI}{\mathcal{I}}
\safemath{\setJ}{\mathcal{J}}
\safemath{\setK}{\mathcal{K}}
\safemath{\setL}{\mathcal{L}}
\safemath{\setM}{\mathcal{M}}
\safemath{\setN}{\mathcal{N}}
\safemath{\setO}{\mathcal{O}}
\safemath{\setP}{\mathcal{P}}
\safemath{\setQ}{\mathcal{Q}}
\safemath{\setR}{\mathcal{R}}
\safemath{\setS}{\mathcal{S}}
\safemath{\setT}{\mathcal{T}}
\safemath{\setU}{\mathcal{U}}
\safemath{\setV}{\mathcal{V}}
\safemath{\setW}{\mathcal{W}}
\safemath{\setX}{\mathcal{X}}
\safemath{\setY}{\mathcal{Y}}
\safemath{\setZ}{\mathcal{Z}}
\safemath{\emptySet}{\varnothing}

\safemath{\colA}{\mathscr{A}}
\safemath{\colB}{\mathscr{B}}
\safemath{\colC}{\mathscr{C}}
\safemath{\colD}{\mathscr{D}}
\safemath{\colE}{\mathscr{E}}
\safemath{\colF}{\mathscr{F}}
\safemath{\colG}{\mathscr{G}}
\safemath{\colH}{\mathscr{H}}
\safemath{\colI}{\mathscr{I}}
\safemath{\colJ}{\mathscr{J}}
\safemath{\colK}{\mathscr{K}}
\safemath{\colL}{\mathscr{L}}
\safemath{\colM}{\mathscr{M}}
\safemath{\colN}{\mathscr{N}}
\safemath{\colO}{\mathscr{O}}
\safemath{\colP}{\mathscr{P}}
\safemath{\colQ}{\mathscr{Q}}
\safemath{\colR}{\mathscr{R}}
\safemath{\colS}{\mathscr{S}}
\safemath{\colT}{\mathscr{T}}
\safemath{\colU}{\mathscr{U}}
\safemath{\colV}{\mathscr{V}}
\safemath{\colW}{\mathscr{W}}
\safemath{\colX}{\mathscr{X}}
\safemath{\colY}{\mathscr{Y}}
\safemath{\colZ}{\mathscr{Z}}

\safemath{\opA}{\mathbb{A}}
\safemath{\opB}{\mathbb{B}}
\safemath{\opC}{\mathbb{C}}
\safemath{\opD}{\mathbb{D}}
\safemath{\opE}{\mathbb{E}}
\safemath{\opF}{\mathbb{F}}
\safemath{\opG}{\mathbb{G}}
\safemath{\opH}{\mathbb{H}}
\safemath{\opI}{\mathbb{I}}
\safemath{\opJ}{\mathbb{J}}
\safemath{\opK}{\mathbb{K}}
\safemath{\opL}{\mathbb{L}}
\safemath{\opM}{\mathbb{M}}
\safemath{\opN}{\mathbb{N}}
\safemath{\opO}{\mathbb{O}}
\safemath{\opP}{\mathbb{P}}
\safemath{\opQ}{\mathbb{Q}}
\safemath{\opR}{\mathbb{R}}
\safemath{\opS}{\mathbb{S}}
\safemath{\opT}{\mathbb{T}}
\safemath{\opU}{\mathbb{U}}
\safemath{\opV}{\mathbb{V}}
\safemath{\opW}{\mathbb{W}}
\safemath{\opX}{\mathbb{X}}
\safemath{\opY}{\mathbb{Y}}
\safemath{\opZ}{\mathbb{Z}}
\safemath{\opZero}{\mathbb{O}}
\safemath{\identityop}{\opI}

\safemath{\veca}{\bma}
\safemath{\vecb}{\bmb}
\safemath{\vecc}{\bmc}
\safemath{\vecd}{\bmd}
\safemath{\vece}{\bme}
\safemath{\vecf}{\bmf}
\safemath{\vecg}{\bmg}
\safemath{\vech}{\bmh}
\safemath{\veci}{\bmi}
\safemath{\vecj}{\bmj}
\safemath{\veck}{\bmk}
\safemath{\vecl}{\bml}
\safemath{\vecm}{\bmm}
\safemath{\vecn}{\bmn}
\safemath{\veco}{\bmo}
\safemath{\vecp}{\bmp}
\safemath{\vecq}{\bmq}
\safemath{\vecr}{\bmr}
\safemath{\vecs}{\bms}
\safemath{\vect}{\bmt}
\safemath{\vecu}{\bmu}
\safemath{\vecv}{\bmv}
\safemath{\vecw}{\bmw}
\safemath{\vecx}{\bmx}
\safemath{\vecy}{\bmy}
\safemath{\vecz}{\bmz}

\safemath{\veczero}{\bmzero}
\safemath{\vecone}{\bmone}
\safemath{\vecxi}{\bmxi}
\safemath{\veclambda}{\bmlambda}
\safemath{\vecmu}{\bmmu}
\safemath{\vectheta}{\bmtheta}
\safemath{\vecphi}{\bmphi}
\safemath{\vecdelta}{\bmdelta}

\safemath{\matA}{\bA}
\safemath{\matB}{\bB}
\safemath{\matC}{\bC}
\safemath{\matD}{\bD}
\safemath{\matE}{\bE}
\safemath{\matF}{\bF}
\safemath{\matG}{\bG}
\safemath{\matH}{\bH}
\safemath{\matI}{\bI}
\safemath{\matJ}{\bJ}
\safemath{\matK}{\bK}
\safemath{\matL}{\bL}
\safemath{\matM}{\bM}
\safemath{\matN}{\bN}
\safemath{\matO}{\bO}
\safemath{\matP}{\bP}
\safemath{\matQ}{\bQ}
\safemath{\matR}{\bR}
\safemath{\matS}{\bS}
\safemath{\matT}{\bT}
\safemath{\matU}{\bU}
\safemath{\matV}{\bV}
\safemath{\matW}{\bW}
\safemath{\matX}{\bX}
\safemath{\matY}{\bY}
\safemath{\matZ}{\bZ}
\safemath{\matzero}{\bmzero}

\safemath{\matDelta}{\bDelta}
\safemath{\matLambda}{\bLambda}
\safemath{\matPhi}{\bPhi}
\safemath{\matSigma}{\bSigma}
\safemath{\matOmega}{\bOmega}
\safemath{\matTheta}{\bTheta}

\safemath{\matidentity}{\matI}
\safemath{\matone}{\matO}

\safemath{\rnda}{A}
\safemath{\rndb}{B}
\safemath{\rndc}{C}
\safemath{\rndd}{D}
\safemath{\rnde}{E}
\safemath{\rndf}{F}
\safemath{\rndg}{G}
\safemath{\rndh}{H}
\safemath{\rndi}{I}
\safemath{\rndj}{J}
\safemath{\rndk}{K}
\safemath{\rndl}{L}
\safemath{\rndm}{M}
\safemath{\rndn}{N}
\safemath{\rndo}{O}
\safemath{\rndp}{P}
\safemath{\rndq}{Q}
\safemath{\rndr}{R}
\safemath{\rnds}{S}
\safemath{\rndt}{T}
\safemath{\rndu}{U}
\safemath{\rndv}{V}
\safemath{\rndw}{W}
\safemath{\rndx}{X}
\safemath{\rndy}{Y}
\safemath{\rndz}{Z}

\safemath{\rveca}{\bimA}
\safemath{\rvecb}{\bimB}
\safemath{\rvecc}{\bimC}
\safemath{\rvecd}{\bimD}
\safemath{\rvece}{\bimE}
\safemath{\rvecf}{\bimF}
\safemath{\rvecg}{\bimG}
\safemath{\rvech}{\bimH}
\safemath{\rveci}{\bimI}
\safemath{\rvecj}{\bimJ}
\safemath{\rveck}{\bimK}
\safemath{\rvecl}{\bimL}
\safemath{\rvecm}{\bimM}
\safemath{\rvecn}{\bimN}
\safemath{\rveco}{\bomO}
\safemath{\rvecp}{\bimP}
\safemath{\rvecq}{\bimQ}
\safemath{\rvecr}{\bimR}
\safemath{\rvecs}{\bimS}
\safemath{\rvect}{\bimT}
\safemath{\rvecu}{\bimU}
\safemath{\rvecv}{\bimV}
\safemath{\rvecw}{\bimW}
\safemath{\rvecx}{\bimX}
\safemath{\rvecy}{\bimY}
\safemath{\rvecz}{\bimZ}

\safemath{\rvecxi}{\bmxi}
\safemath{\rveclambda}{\bmlambda}
\safemath{\rvecmu}{\bmmu}
\safemath{\rvectheta}{\bmtheta}
\safemath{\rvecphi}{\bmphi}

\safemath{\rmatA}{\bimA}
\safemath{\rmatB}{\bimB}
\safemath{\rmatC}{\bimC}
\safemath{\rmatD}{\bimD}
\safemath{\rmatE}{\bimE}
\safemath{\rmatF}{\bimF}
\safemath{\rmatG}{\bimG}
\safemath{\rmatH}{\bimH}
\safemath{\rmatI}{\bimI}
\safemath{\rmatJ}{\bimJ}
\safemath{\rmatK}{\bimK}
\safemath{\rmatL}{\bimL}
\safemath{\rmatM}{\bimM}
\safemath{\rmatN}{\bimN}
\safemath{\rmatO}{\bimO}
\safemath{\rmatP}{\bimP}
\safemath{\rmatQ}{\bimQ}
\safemath{\rmatR}{\bimR}
\safemath{\rmatS}{\bimS}
\safemath{\rmatT}{\bimT}
\safemath{\rmatU}{\bimU}
\safemath{\rmatV}{\bimV}
\safemath{\rmatW}{\bimW}
\safemath{\rmatX}{\bimX}
\safemath{\rmatY}{\bimY}
\safemath{\rmatZ}{\bimZ}

\safemath{\rmatDelta}{\bimDelta}
\safemath{\rmatLambda}{\bimLambda}
\safemath{\rmatPhi}{\bimPhi}
\safemath{\rmatSigma}{\bimSigma}
\safemath{\rmatOmega}{\bimOmega}
\safemath{\rmatTheta}{\bimTheta}

\usepackage{amssymb}
\usepackage{amsfonts}
\usepackage{mathrsfs}
\usepackage{xspace}
\usepackage{bm}
\usepackage{fancyref}
\usepackage{textcomp}

\usepackage{multirow}
\usepackage{stmaryrd}

\newenvironment{textbmatrix}{	\setlength{\arraycolsep}{2.5pt}%
								\big[\begin{matrix}}{\end{matrix}\big]%
								\raisebox{0.08ex}{\vphantom{M}}}

\def\be{\begin{equation}}
\def\ee{\end{equation}}
\def\een{\nonumber \end{equation}}
\def\mat{\begin{bmatrix}}
\def\emat{\end{bmatrix}}
\def\btm{\begin{textbmatrix}}
\def\etm{\end{textbmatrix}}

\def\ba#1\ea{\begin{align}#1\end{align}}
\def\bas#1\eas{\begin{align*}#1\end{align*}}
\def\bs#1\es{\begin{split}#1\end{split}}
\def\bg#1\eg{\begin{gather}#1\end{gather}}
\def\bml#1\eml{\begin{multline}#1\end{multline}}
\def\bi#1\ei{\begin{itemize}#1\end{itemize}}

\newcommand{\lefto}{\mathopen{}\left}

\DeclareMathOperator*{\argmin}{arg\;min}		
\DeclareMathOperator*{\argmax}{arg\;max}		
\DeclareMathOperator{\Exop}{\opE}			

\newcommand{\Ex}[2]{\ensuremath{\Exop_{#1}\lefto[#2\right]}} 	

\newcommand{\tp}[1]{\ensuremath{#1^{\text{T}}}} 		
\newcommand{\herm}[1]{\ensuremath{#1^{\text{H}}}} 	
\newcommand{\inv}[1]{\ensuremath{#1^{-1}}} 	
\newcommand{\pinv}[1]{\ensuremath{#1^{\dagger}}} 	

\safemath{\dirac}{\delta}					
\safemath{\krond}{\dirac}					

\safemath{\upto}{\uparrow}
\safemath{\downto}{\downarrow}
\safemath{\iu}{j}							
\safemath{\ev}{\lambda}						
\safemath{\hilseqspace}{l^{2}}				
\newcommand{\banachfunspace}[1]{\setL^{#1}}	
\safemath{\hilfunspace}{\banachfunspace{2}}	

\safemath{\SNR}{\textit{SNR}} 				
\safemath{\PAR}{\textit{PAR}} 				
\safemath{\No}{N_0}							
\safemath{\Es}{E_s}							
\safemath{\Eb}{E_b}							
\safemath{\EbNo}{\frac{\Eb}{\No}}
\safemath{\EsNo}{\frac{\Es}{\No}}

\DeclareMathOperator{\CHop}{\ensuremath{\opH}} 
\safemath{\tvir}{\rndh_{\CHop}}				
\safemath{\tvtf}{\rndl_{\CHop}}				
\safemath{\spf}{\rnds_{\CHop}}				
\safemath{\bff}{H_{\CHop}}					

\safemath{\ircf}{r_{h}}						
\safemath{\tftvcf}{r_{s}}					
\safemath{\tfcf}{r_{l}}						
\safemath{\bfcf}{r_{H}}						

\safemath{\tcorr}{c_h}						
\safemath{\scf}{c_{s}}						
\safemath{\tfcorr}{c_{l}}					
\safemath{\fcorr}{c_{H}}						

\safemath{\mi}{I}							
\safemath{\capacity}{C}						

\safemath{\normal}{\mathcal{N}}			
\safemath{\jpg}{\mathcal{CN}}			
\safemath{\mchain}{\leftrightarrow}		

\safemath{\dB}{\,\mathrm{dB}}
\safemath{\dBm}{\,\mathrm{dBm}}
\safemath{\Hz}{\,\mathrm{Hz}}
\safemath{\kHz}{\,\mathrm{kHz}}
\safemath{\MHz}{\,\mathrm{MHz}}
\safemath{\GHz}{\,\mathrm{GHz}}
\safemath{\s}{\,\mathrm{s}}
\safemath{\ms}{\,\mathrm{ms}}
\safemath{\mus}{\,\mathrm{\text{\textmu}s}}
\safemath{\ns}{\,\mathrm{ns}}
\safemath{\ps}{\,\mathrm{ps}}
\safemath{\meter}{\,\mathrm{m}}
\safemath{\mm}{\,\mathrm{mm}}
\safemath{\cm}{\,\mathrm{cm}}
\safemath{\m}{\,\mathrm{m}}
\safemath{\W}{\,\mathrm{W}}
\safemath{\mW}{\, \mathrm{mW}}
\safemath{\J}{\,\mathrm{J}}
\safemath{\K}{\,\mathrm{K}}
\safemath{\bit}{\,\mathrm{bit}}
\safemath{\nat}{\,\mathrm{nat}}

\safemath{\define}{\triangleq}			

\safemath{\equivalent}{\sim}
\safemath{\distas}{\sim}					
\safemath{\sdiff}{\Delta}				

\safemath{\reals}{\mathbb{R}}
\safemath{\positivereals}{\reals_{+}}
\safemath{\integers}{\mathbb{Z}}
\safemath{\posint}{\integers_{+}}
\safemath{\naturals}{\mathbb{N}}
\safemath{\posnaturals}{\naturals_{+}}
\safemath{\complexset}{\mathbb{C}}
\safemath{\rationals}{\mathbb{Q}}

\newcommand*{\fancyrefapplabelprefix}{app}		
\newcommand*{\fancyrefthmlabelprefix}{thm}		
\newcommand*{\fancyreflemlabelprefix}{lem}		
\newcommand*{\fancyrefcorlabelprefix}{cor}		
\newcommand*{\fancyrefdeflabelprefix}{def}		
\newcommand*{\fancyrefproplabelprefix}{prop}		
\newcommand*{\fancyrefexmpllabelprefix}{exmpl}
\newcommand*{\fancyrefalglabelprefix}{alg}		
\newcommand*{\fancyreftbllabelprefix}{tbl}		

\frefformat{vario}{\fancyrefseclabelprefix}{Section~#1}
\frefformat{vario}{\fancyrefthmlabelprefix}{Theorem~#1}
\frefformat{vario}{\fancyreftbllabelprefix}{Table~#1}
\frefformat{vario}{\fancyreflemlabelprefix}{Lemma~#1}
\frefformat{vario}{\fancyrefcorlabelprefix}{Corollary~#1}
\frefformat{vario}{\fancyrefdeflabelprefix}{Definition~#1}
\frefformat{vario}{\fancyreffiglabelprefix}{Fig.~#1}
\frefformat{vario}{\fancyrefapplabelprefix}{Appendix~#1}
\frefformat{vario}{\fancyrefeqlabelprefix}{(#1)}
\frefformat{vario}{\fancyrefproplabelprefix}{Proposition~#1}
\frefformat{vario}{\fancyrefexmpllabelprefix}{Example~#1}
\frefformat{vario}{\fancyrefalglabelprefix}{Algorithm~#1}

 \newtheorem{thm}{Theorem}
 
 \newtheorem{defi}{Definition}
 
 \newtheorem{exmpl}{Example}
 \newtheorem{remark}{Remark}
 \newtheorem*{remark*}{Remark}

\safemath{\dictab}{[\,\dicta\,\,\dictb\,]}

\safemath{\ysig}{\bmy}
\safemath{\ysighat}{\hat{\ysig}}
\safemath{\ysigdim}{M}
\safemath{\xsig}{\bmx}
\safemath{\xsigdim}{N}
\safemath{\nx}{n_x}
\safemath{\zsig}{\bmz}
\safemath{\zsigdim}{\ysigdim}
\safemath{\rsig}{\bmr}
\safemath{\Adict}{\bA}
\safemath{\Adicttilde}{\widetilde{\Adict}}
\safemath{\Adictdim}{\outputdim\times\xsigdim}
\safemath{\avec}{\bma}
\safemath{\avectilde}{\tilde{\avec}}
\safemath{\Bdict}{\bB}
\safemath{\Bdicttilde}{\widetilde{\Bdict}}
\safemath{\Cdict}{\bC}
\safemath{\cvec}{\bmc}
\safemath{\Ddict}{\bD}
\safemath{\Ddictdim}{\ysigdim\times\xsigdim}
\safemath{\dvec}{\bmd}
\safemath{\Ddicttilde}{\widetilde{\bD}}
\safemath{\Bonb}{\bB}
\safemath{\bvec}{\bmb}
\safemath{\Bonbdim}{\ysigdim\times\ysigdim}
\safemath{\noise}{\bmn}
\safemath{\noisedim}{\ysigim}
\safemath{\err}{\bme}
\safemath{\errdim}{\ysigdim}
\safemath{\errset}{\setE}
\safemath{\nerr}{n_e}
\safemath{\delop}{\bP_\errset}
\safemath{\delopc}{\bP_{{\errset}^c}}

\safemath{\cplxi}{\imath}
\safemath{\cplxj}{\jmath}

\safemath{\dict}{\matD}
\safemath{\inputdim}{N}		
\safemath{\outputdim}{M}		
\safemath{\sparsity}{S}	
\safemath{\inputdimA}{{N_a}}	
\safemath{\inputdimB}{{N_b}}	
\safemath{\elemA}{{n_a}}	
\safemath{\elemB}{{n_b}}	
\safemath{\resA}{\matR_a}	
\safemath{\resB}{\matR_b}	
\safemath{\subD}{\matS} 
\safemath{\subA}{\matS_a} 
\safemath{\subB}{\matS_b} 
\safemath{\dicta}{\matA} 	
\safemath{\dictb}{\matB} 	
\safemath{\hollowS}{H}
\safemath{\hollowA}{H_a}
\safemath{\hollowB}{H_b}
\safemath{\cross}{Z}
\safemath{\coh}{\mu_d}			
\safemath{\coha}{\mu_a}			
\safemath{\cohb}{\mu_b}			
\safemath{\mubs}{\nu}	
\safemath{\cohm}{\mu_m} 
\safemath{\dictset}{\setD}	
\safemath{\dictsetp}{\dictset(\coh,\coha,\cohb)}	
\safemath{\dictsetgen}{\dictset_\text{gen}}
\safemath{\dictsetgenp}{\dictsetgen(\coh)}
\safemath{\dictsetonb}{\dictset_\text{onb}}
\safemath{\dictsetonbp}{\dictsetonb(\coh)}

\safemath{\leftside}{U}
\safemath{\rightsideA}{R_a}
\safemath{\rightsideB}{R_b}

\safemath{\indexS}{\setI_S} 

\safemath{\na}{n_a}			
\safemath{\nb}{n_b}			
\safemath{\coeffa}{p_i}	
\safemath{\coeffb}{q_j}	
\safemath{\seta}{\setP}		
\safemath{\setb}{\setQ}     
\safemath{\setw}{\setW}	
\safemath{\setz}{\setZ}	
\safemath{\cola}{\veca}		
\safemath{\colb}{\vecb}		
\safemath{\cold}{\vecd}		
\safemath{\inputvec}{\vecx} 	
\safemath{\error}{\vece}	
\safemath{\noiseout}{\vecz} 	
\safemath{\inputvecel}{x}
\safemath{\inputveca}{\vecx_a}
\safemath{\inputvecb}{\vecx_b}
\safemath{\outputvec}{\vecy}	
\safemath{\lambdamin}{\lambda_{\mathrm{min}}}

\safemath{\elltwo}{\ell_2}
\safemath{\ellone}{\ell_1}
\safemath{\ellzero}{\ell_0}
\safemath{\ellinf}{\ell_\infty}
\safemath{\ellinftilde}{\ell_{\widetilde\infty}}
\safemath{\licard}{Z(\coh,\coha,\cohb)}
\safemath{\xsol}{\hat{x}}
\safemath{\xbord}{x_b}		
\safemath{\xstat}{x_s}		
\safemath{\xstatLone}{\tilde{x}_s}
\safemath{\order}{\mathcal{O}} 
\safemath{\scales}{\Theta} 
\safemath{\ones}{\mathbf{1}} 
\safemath{\zeroes}{\mathbf{0}} 
\safemath{\thlone}{\kappa(\coh,\cohb)} 
\safemath{\constoneA}{\delta} 
\safemath{\constoneB}{\epsilon} 
\safemath{\nlarge}{L}				   
\safemath{\sumlarge}{S_\nlarge}
\safemath{\maxlarger}{P_\nlarge}	   
\safemath{\Pzero}{\textrm{P0}}	
\safemath{\Pone}{\textrm{P1}}
\safemath{\vecfir}{\vecw}			 
\safemath{\vecsec}{\vecz}
\safemath{\elvecfir}{w}              
\safemath{\elvecsec}{z}				 
\safemath{\nlargefir}{n}
\safemath{\normout}{\gamma}
\safemath{\auxfun}{h}
\safemath{\supp}{\textrm{supp}}

\safemath{\indexa}{\ell}
\safemath{\indexb}{r}
\safemath{\indexc}{i}
\safemath{\indexd}{j}

\safemath{\project}{P}

\usepackage{framed}

\IEEEoverridecommandlockouts
\allowdisplaybreaks 

\newcommand{\blue}[1]{\textcolor{black}{#1}}

\safemath{\Hj}{\bmj}
\safemath{\bsj}{\bmw}
\safemath{\sj}{w}
\safemath{\Ej}{E_w}
\safemath{\proxg}{\text{prox}_g}
\safemath{\pma}{\text{pma}_\setS}
\safemath{\rE}{\rho_{\textsf{\tiny{E}}}}
\safemath{\rP}{\rho_{\textsf{\tiny{P}}}}

\begin{document}
\bstctlcite{IEEEexample:BSTcontrol} 

\title{Mitigating Smart Jammers in Multi-User MIMO}

\author{\IEEEauthorblockN{Gian Marti, Torben K\"olle, and Christoph Studer}
\thanks{A short version of this paper has been presented at IEEE ICC 2022~\cite{marti2022smart}.
Besides providing a more complete discussion in general, 
the present paper extends the work in \cite{marti2022smart} by providing theoretical underpinnings of our approach, 
by providing the SO-MAED algorithm which improves upon the MAED algorithm from \cite{marti2022smart},
and by extending the evaluation to higher-order constellations (instead of only QPSK)
as well as to ray-traced mmWave channels (instead of only Rayleigh fading).}
\thanks{The work of CS was supported in part by ComSenTer, one of six centers in JUMP, a SRC program sponsored by DARPA, 
and in part by the U.S.\ National Science Foundation (NSF) under grants CNS-1717559 and ECCS-1824379. The work of GM and CS was supported in part by an ETH Research~Grant.}
\thanks{The authors thank H. Song for discussions on joint channel estimation and data detection, 
\blue{and S. Taner for helpful comments on the manuscript.}}
\thanks{G. Marti, T. K\"olle, and C. Studer are with the Department of Information Technology
and Electrical Engineering, ETH Zurich, Switzerland; e-mail: gimarti@ethz.ch, tkoelle@student.ethz.ch, and studer@ethz.ch}
}

\maketitle

\begin{abstract}
Wireless systems must be resilient to jamming attacks. 
Existing mitigation methods based on multi-antenna processing require knowledge of the jammer's  transmit characteristics
that may be difficult to acquire, especially for smart jammers that evade mitigation by transmitting only at specific instants.
We propose a novel method to mitigate smart jamming attacks on the massive multi-user multiple-input multiple-output \mbox{(MU-MIMO)} uplink
which does not require the jammer to be active at any specific instant. 
By formulating an optimization problem that unifies jammer estimation and mitigation, channel estimation, and data detection, 
we exploit that a jammer cannot change its subspace within a coherence interval.
Theoretical results for our problem formulation show that its solution is guaranteed to recover the users' data symbols under certain conditions.
We develop two efficient iterative algorithms for approximately solving the proposed problem formulation:  
MAED, a parameter-free algorithm which uses 
forward-backward splitting with a box symbol prior, and SO-MAED, which replaces the prior of MAED with soft-output symbol estimates that exploit the discrete 
transmit constellation and which uses deep unfolding to optimize algorithm parameters. 
We use simulations to demonstrate that the proposed algorithms effectively mitigate a wide range of smart 
jammers without a priori knowledge about the attack type. 
\end{abstract}
\section{Introduction}
\IEEEPARstart{J}{amming} attacks pose a serious threat to the continuous operability of wireless communication systems \cite{economist2021satellite, topgun}. 
Effective methods to mitigate such attacks are of paramount importance
as wireless systems become increasingly critical to modern infrastructure~\cite{popovski2014ultra, pirayesh2022jamming}.
\blue{In the massive multi-user multiple-input multiple-output (MU-MIMO) uplink, effective jammer mitigation becomes possible by the strong asymmetry in the number of antennas between the basestation (BS), which has many antennas, and a mobile jamming device, which typically has one or few antennas.}
One possibility \blue{for jammer mitigation}, for instance, is to project the receive signals on the subspace 
orthogonal to the jammer's channel~\cite{marti2021snips,yan2016jamming}. 
Unfortunately, such methods require accurate knowledge of the jammer's channel.
If a jammer transmits permanently and with a static signature (often called barrage jamming), the~BS~can estimate its channel, for instance during a dedicated period in which the user equipments (UEs) do not transmit~\cite{marti2021snips} or in which they transmit predefined symbols~\cite{yan2016jamming}.
In contrast to barrage jamming, however, a smart jammer might jam the system only at specific time instants, such as when the UEs are transmitting data symbols,
and thereby prevent the BS from estimating the jammer's channel using simple estimation~algorithms.

\subsection{State of the Art}

\begin{figure}[tp]
\centering
\subfigure[Existing methods separate jammer estimation (JEST) and channel~estimation 
(CHEST) from the jammer-resilient data detection (DET). They~are ineffective against 
jammers that jam the data phase but not the training~phase.]
{
\includegraphics[width=0.94\columnwidth]{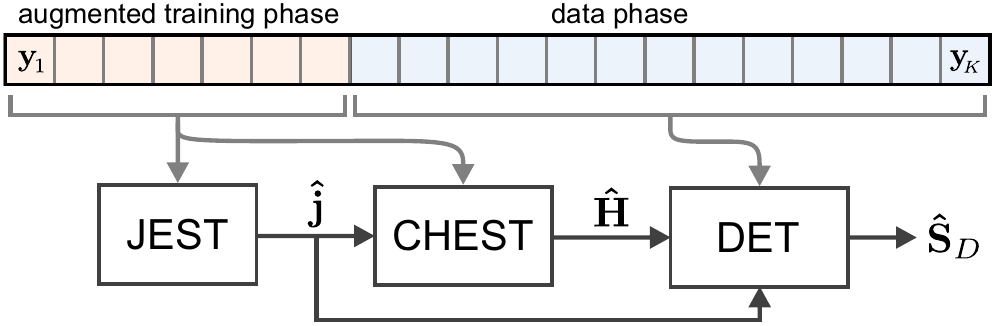}
\label{fig:traditional}
}
\newline
\subfigure[Our method unifies jammer estimation and mitigation, channel estimation, 
and data detection to deal with jammers regardless of their activity~pattern.]
{
\includegraphics[width=0.94\columnwidth]{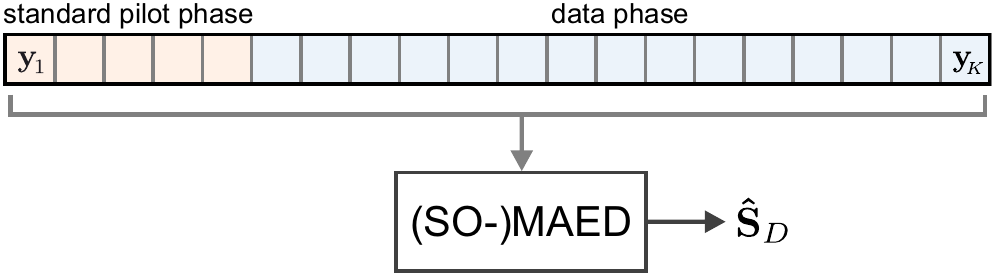}
\label{fig:maed}
}
\caption{
The approach to jammer mitigation taken by existing methods (a) compared to the proposed method (b).
In the figure, $\bmy_1,\dots,\bmy_K$ are the receive signals, 
and $\hat\Hj, \hat\bH$, and $\hat\bS_D$ are the estimates of the jammer channel, the UE channel matrix, and the UE \blue{data} symbols, respectively.
}
\vspace{-2mm}
\label{fig:maed_vs_trad}
\end{figure}

Multi-antenna wireless systems offer the unique potential to effectively mitigate jamming attacks.
Consequently, a variety of multi-antenna methods have been proposed for the mitigation of jamming attacks in MIMO systems
\cite{pirayesh2022jamming, marti2021snips, shen14a, hoang2021suppression, yan2016jamming,zeng2017enabling, vinogradova16a, do18a, akhlaghpasand20a, akhlaghpasand20b, marti2021hybrid, wan2022robust, darsena2022anti}.
Common to all of them~is the assumption---in one way or other---that information about the jammer's transmit characteristics
(e.g., the jammer's channel, or the covariance matrix between the UE transmit signals and the jammed receive signals)
can be estimated using some specific subset of the receive samples.\footnote{The method of \cite{vinogradova16a} is to some extent an exception as it estimates the~UEs' subspace and projects the receive signals thereon. This method, however, dist-inguishes the UEs' from the jammer's subspace based on the receive power, thereby presuming that the UEs and the jammer transmit with different power.}
\fref{fig:traditional}~illustrates the approach of such methods: 
The data phase is preceded by an augmented 
training phase in which the jammer's transmit characteristics as well as the channel matrix are estimated.
This augmented training phase may (i)~complement a traditional pilot phase with a dedicated period during which the UEs do not transmit in order to 
enable jammer estimation (e.g., \cite{marti2021snips, shen14a, hoang2021suppression}) or (ii) consist of an extended pilot phase so that there exist pilot sequences that are unused by
the UEs and on whose span the receive signals can be projected to estimate the jammer subspace~\mbox{(e.g., \cite{do18a, akhlaghpasand20a, akhlaghpasand20b}).}
The estimated jammer characteristics are then used to perform jammer-mitigating data detection. 
Such an approach succeeds in the case of barrage jammers, but is unreliable for estimating the propagation characteristics of smart jammers, see \fref{sec:example}: 
A smart jammer can evade estimation and thus circumvent mitigation by not transmitting during the training phase, for instance because it is aware of the defense mechanism
or simply because it jams in short bursts only. 
For this reason, our proposed method does not estimate the jammer channel based on a dedicated training phase, 
but instead utilizes the entire transmission period and unifies jammer estimation and mitigation, channel estimation and data detection; see \fref{fig:maed}.

Many studies have already shown how smart jammers can disrupt wireless communication systems by
targeting only specific parts
of the wireless transmission process \cite{miller2010subverting, miller2011vulnerabilities, clancy2011efficient, sodagari2012efficient, 
lichtman2013vulnerability, lichtman20185g, lichtman2016lte, girke2019towards,lapan2012jamming} 
instead of using barrage jamming.
Jammers that target only the pilot phase have received considerable attention 
\cite{miller2010subverting,miller2011vulnerabilities,clancy2011efficient,sodagari2012efficient,lichtman2013vulnerability}, 
as such attacks can be more energy-efficient than barrage jamming in disrupting communication systems that do not 
defend themselves against jammers~\cite{clancy2011efficient,sodagari2012efficient,lichtman2013vulnerability}.
However, if a jammer is active during the pilot phase, then a BS that \emph{does} defend itself against attacks
can estimate the jammer's channel by exploiting knowledge of the UE transmit symbols during the pilot phase, for instance with the aid of unused pilot sequences~\cite{do18a, akhlaghpasand20a, akhlaghpasand20b}.
To disable such jammer-mitigating communication systems, a smart jammer might thus refrain from jamming the pilot phase and only target 
the data phase, even if such jamming attacks have not received much attention so far 
\cite{lichtman2016lte, girke2019towards}.
Other threat models that have been analyzed include attacks on control 
channels \cite{lichtman2013vulnerability, lichtman2016lte, lichtman20185g}, the beam alignment procedure \cite{darsena2022anti}, 
or the time synchronization phase~\cite{lapan2012jamming},\blue{\cite{el2017lte}}, but this paper will not consider such protocol or control channel attacking schemes. 

\subsection{Contributions}
To mitigate smart jammers \blue{in the massive MU-MIMO uplink,} 
we propose a novel approach that does not depend on the jammer being active during 
\textit{any} specific period.
Leveraging the fact that a jammer cannot change its subspace instantaneously,
we utilize a problem formulation which unifies jammer estimation and mitigation, 
channel estimation, and data detection, instead of dealing with these tasks independently (cf.~\fref{fig:maed}).
We support the soundness of the proposed optimization problem by proving that its global minimum is unique and recovers the transmitted data symbols, given that certain conditions are satisfied. 
By building on techniques for joint channel estimation and data detection
\cite{vikalo2006efficient, xu2008exact, kofidis2017joint, castaneda2018vlsi, yilmaz2019channel, he2020model, song2021soft},
we then develop two efficient iterative algorithms for approximately solving the optimization 
problem. The first algorithm is called MAED (short for MitigAtion, Estimation, and Detection) and 
solves the problem approximately using forward-backward splitting (FBS) \cite{goldstein16a}. 
The second algorithm is called SO-MAED (short for Soft-Output MAED) and extends MAED with a more informative 
prior on the data symbols to produce soft symbol estimates. SO-MAED relies on deep unfolding to optimize its parameters {\cite{song2021soft, hershey2014deep, balatsoukas2019deep, goutay2020deep, monga2021algorithm}.
We use simulations with different propagation models to demonstrate that MAED and SO-MAED effectively mitigate a wide variety of 
na\"ive and smart  jamming attacks without requiring any knowledge about the attack type.

\subsection{Notation}
Matrices and column vectors are represented by boldface uppercase and lowercase letters, respectively.
For a matrix~$\bA$, \blue{the conjugate is $\bA^{\!\ast}$,} the transpose is $\tp{\bA}$, the conjugate~transpose is $\herm{\bA}$, 
the Moore-Penrose pseudoinverse is $\pinv{\bA}$,
the entry in the $\ell$th row and $k$th column is $[\bA]_{\ell,k}$, 
the $k$th column is $\bma_k$,
the submatrix consisting of the columns from $n$ through $m$ is $\bA_{[n:m]}$,
and the Frobenius norm is $\| \bA \|_F$.
The $N\!\times\!N$ identity matrix is $\bI_N$.
For a vector~$\bma$, the $\ell_2$-norm is $\|\bma\|_2$, the real part is $\Re\{\bma\}$, the imaginary part is $\Im\{\bma\}$, 
and the span is $\textit{span}(\bma)$.
\blue{For vectors $\bma, \bmb$, we define $[\tp{\bma};\tp{\bmb}]\triangleq \tp{[\bma, \bmb]}$.}
Expectation with respect to a random vector~$\bmx$ is denoted by \Ex{\bmx}{\cdot}.
We define $i^2=-1$. 
The complex $n$-hypersphere of radius $r$ is denoted by $\mathbb{S}_r^n$,
and~$[n:m]$ are the integers from $n$ through~$m$.

\section{System Setup}\label{sec:setup}
\subsection{\blue{Transmission Model}}
We consider the uplink of a massive MU-MIMO system in which $U$ single-antenna UEs transmit data to 
a $B$ antenna BS in the presence of a single-antenna jammer.
The channels are assumed to be frequency flat and block-fading with coherence time $K=T+D$.
The first $T$ time slots are used to 
transmit pilot symbols; the remaining $D$ time slots are used to transmit data symbols.
The UE transmit matrix is $\bS = [\bS_T,\bS_D]$, where $\bS_T\in \opC^{U\times T}$
and $\bS_D\in\setS^{U\times D}$ contain the pilots~and the \blue{data} symbols, respectively. 
\blue{The data symbols $\bS_D$ are drawn i.i.d. uniformly from a constellation $\setS$,}
which is normalized to unit average symbol energy.
We assume that the jammer does not prevent the UEs and the BS from establishing synchronization,
which allows us to use the discrete-time input-output relation
\begin{align}
	\bY = \bH\bS + \Hj\tp{\bsj} + \bN. \label{eq:io}
\end{align}
Here, $\bY\in\opC^{B\times K}$ is the BS receive matrix that contains the \mbox{$B$-dimensional} receive vectors over all $K$ time slots, 
\mbox{$\bH\in\opC^{B\times U}$} models the channel between the UEs and the~BS,
$\Hj\in\opC^B$ models the channel between the jammer and the~BS, 
\blue{$\tp{\bsj}=[\tp{\bsj_T},\tp{\bsj_D}]\in\opC^K$} contains the jammer transmit symbols over all $K$ time slots, 
and $\bN\in\opC^{B\times K}$ models thermal noise consisting of independently and identically distributed (i.i.d.) circularly-symmetric complex Gaussian entries with variance~$N_0$.
Unless stated otherwise, we assume that the jammer's transmit symbols $\bsj$ are independent of $\bS$. 
No other assumptions about the distribution of $\bsj$ are made;
in particular, we do not assume that these entries are~i.i.d.

In what follows, we use plain symbols for the true channels and transmit signals, variables with a tilde for optimization variables, and quantities with a hat for (approximate) solutions to optimization problems, e.g., $\hat\bS_D$ is the estimate of the UE \blue{data} symbol matrix~$\bS_D$ as determined by solving an optimization problem with respect to $\tilde\bS_D$.

\subsection{\blue{Model Limitations}} \label{sec:limitations}
\blue{We now point out---and discuss the relevance of---a number of limitations of our transmission model. 
\newline \indent
Our model only considers  single-antenna UEs, while multi-antenna UEs and point-to-point (p2p) MIMO are
excluded. In principle, our method could also be combined with multi-antenna
UEs or p2p MIMO, as long as spatial multiplexing is used. This would simply change the
transmission model\footnote{\blue{The statistics of $\bH$ would 
also change compared to the  single-antenna UE case, but our methods do not depend on particular 
channel statistics.}} in \eqref{eq:io} to $\bY = \bH\bF\bS + \Hj\tp{\bsj} + \bN$, where $\bF$
is a transmit beamforming matrix which is either block-diagonal (in the case of multi-antenna UEs) 
or dense (in the case of p2p MIMO). Such a model would raise the question of how to choose the transmit
beamformer(s) $\bF$, and how to obtain the necessary channel state information at the transmitter(s)
in the presence jamming. We leave these issues for future work.
\newline \indent
Similarly, we only consider single-antenna jammers. However, the ideas that underlie our
methods can also be extended to the mitigation of multi-antenna jammers. We consider the 
mitigation of multi-antenna jammers with methods similar---but not identical---to the ones in this paper
in \cite{marti2022joint}.
\newline \indent
Another limitation pertains to our use of a block-fading channel model. Real-world channels do 
not stay constant for a fixed amount of time and then change abruptly. However, our method does
not depend on how the channel changes between coherence blocks, but only on the channel staying
(approximately) constant for a certain period of time, which is a reasonable assumption in practice. 
In real-world channels which change continuously even between coherence intervals, channel knowledge
from previous coherence intervals could potentially be used to find effective initializers for our algorithms. 
We defer such investigations to future work. 
\newline \indent
We also, for the most part, assume independence between the jammer's transmit symbols $\bmw$ and the 
UEs' transmit signals $\bS$, which comprise the pilots $\bS_T$ and the data symbols $\bS_D$. 
This is motivated by the reasonable assumption that the UEs' transmit data are
\emph{a priori} unknown to anyone except themselves. 
The jammer's time-$k$ transmit symbol $w_k$ can thus not depend on the time-$k$ UE data vector $\bms_k$.
In principle, the jammer could try to detect the UE data symbols to make $w_k$ 
dependent on $\bms_1,\dots,\bms_{k-L}$,
for some processing latency $L\geq 1$. However, such ``full-duplex'' jamming would be extremely difficult
to implement \cite{sabharwal2014band}. Also, delayed dependencies (such as replaying the signal 
of some UE with a delay) would have no bearing on the performance of our method, which
does not ``mix'' the receive signals from different time indexes in processing.
The assumption that the jammer transmit symbols do not depend on the pilots $\bS_T$ is thus
reasonable \emph{as long as randomized pilots are used} (cf. \fref{sec:theory}), 
but not necessarily when the pilots are deterministic and known to the jammer
(cf. \fref{sec:results:eclipsed}).
\newline \indent
The final limitation is our assumption of perfect synchronization between the UEs and the BS. 
This is not an innocent assumption, as jammers can inhibit synchronization \cite{lapan2012jamming}. 
However, we consider the question of how to synchronize in the presence of jamming as a separate
research problem that is outside the scope of this paper
and for which we refer~to~\cite{el2017lte}.}

\begin{figure}[tp]
\centering
\!\!
\subfigure[mitigation of a barrage jammer]{
\includegraphics[height=3.85cm]{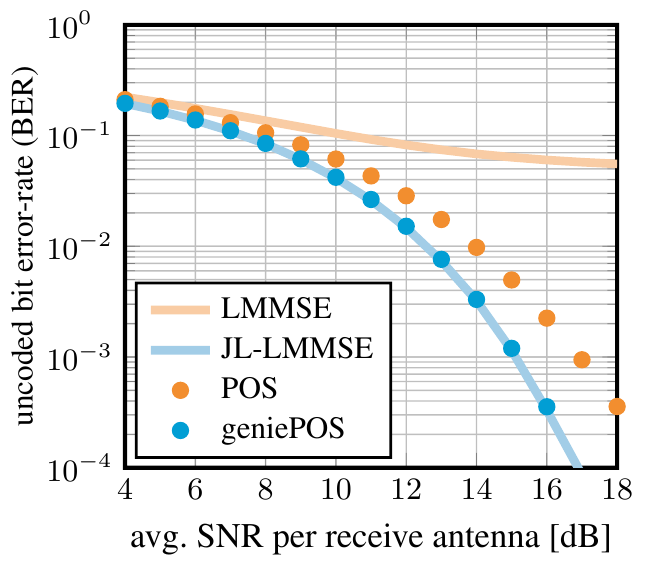}
\label{fig:example:success}
}\!\!\!
\subfigure[failed mitigation of a smart jammer]{
\includegraphics[height=3.85cm]{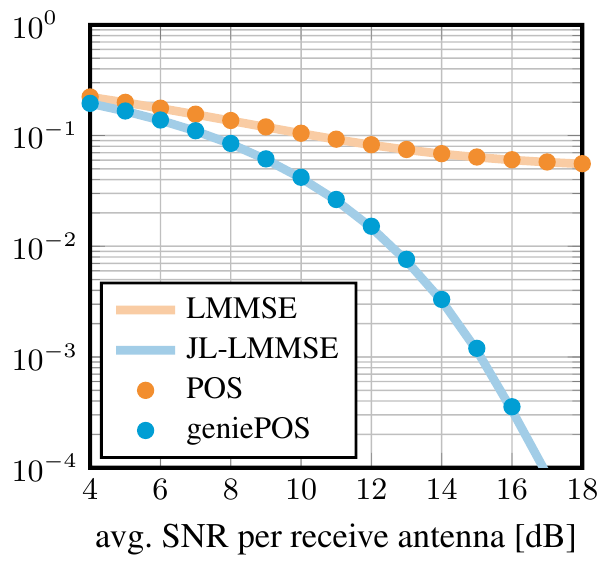}
\label{fig:example:fail}
}\!\!
\caption{\blue{An} example that illustrates how methods that estimate the jammer's channel based on 
a subset of samples fail when facing a smart jammer.}
\label{fig:example}
\end{figure}

\section{Motivating Example} \label{sec:example}

\blue{To understand the challenge posed by smart jammers as opposed to barrage jammers, we}
start by considering the motivating example of \fref{fig:example}, 
which shows uncoded bit error-rates (BERs) of different receivers \blue{(LMMSE, JL-LMMSE, geniePOS, POS)} for an i.i.d. Rayleigh 
fading MU-MIMO system with \mbox{$B=128$} BS antennas and $U=32$ UEs that transmit 16-QAM symbols under~a jamming attack.
In \fref{fig:example:success} the system is attacked by a barrage jammer that
transmits i.i.d. Gaussian symbols and whose 
receive power exceeds that of the average UE by 30\,dB. 
\blue{The different receivers operate as follows:}
\subsubsection{\blue{LMMSE}} 
\blue{This receiver estimates the channel matrix 
using orthogonal pilots with a least squares (LS) estimator followed by a linear minimum mean square error (LMMSE) detector.}

\subsubsection{\blue{JL-LMMSE}} 
\blue{This receiver works identical to the LMMSE receiver but operates in 
a jammerless (``JL'') system.}
\subsubsection{\blue{geniePOS}}
\blue{This} receiver serves as a baseline and is furnished with ground-truth knowledge of the jammer channel~$\Hj$. It  
nulls the jammer by orthogonally projecting the receive signals on the orthogonal 
complement \blue{(``POS'' is short for Projection onto the Orthogonal Subspace)} of $\textit{span}(\Hj)$ using the matrix $\bP_\Hj = \bI_B - \Hj\pinv{\Hj}$\blue{\cite[Sec. 2.6.1]{GV96}}, 
where $\pinv{\Hj}=\herm{\Hj}/\|\Hj\|_2^2$,~as
\begin{align}
	\bP_\Hj\bY 
	&= \bP_\Hj\,\bH\bS + \bP_\Hj\,\Hj\tp{\bsj} + \bP_{\Hj}\,\bN \label{eq:pos} \\
	&= \bP_\Hj\,\bH\bS + \bP_\Hj\,\bN,
\end{align}
since $\bP_\Hj\,\Hj=\mathbf{0}$.
The result is an effective jammerless system with receive signal $\bY_{\bP} = \bP_\Hj \bY$, 
effective channel matrix \mbox{$\bH_{\bP} = \bP_\Hj\bH$}, and (colored) noise 
$\bN_\bP = \bP_\Hj \bN \sim \setC\setN(\mathbf{0},\No\bP_\Hj)$.
Finally, geniePOS performs LS channel estimation and subsequent LMMSE data detection in this 
projected system  \cite{marti2021snips}. 
\subsubsection{\blue{POS}}
\blue{This} receiver works analogously to geniePOS, except that it is not furnished with ground-truth 
knowledge of the jammer channel---instead, this \blue{receiver} estimates the jammer subspace $\Hj/\|\Hj\|_2$
based on ten receive samples in which the UEs do not transmit and only the jammer is active. 
If the matrix received in that period is denoted by $\bY_\text{J}$, then the jammer subspace is estimated
as the left-singular vector of the largest singular value of $\bY_\text{J}$.

\fref{fig:example:success} shows that geniePOS effectively mitigates the jammer, achieving a performance
virtually identical to that of the jammer-free \mbox{JL-LMMSE} receiver. Indeed, geniePOS nulls the jammer 
perfectly, so that the only performance loss comes from the loss of one degree-of-freedom
in the receive signal. POS is not as effective,~since it nulls the jammer only imperfectly due
to its noisy estimate of the jammer subspace. However, this method still mitigates the jammer with
a loss of less than 2\,dB in SNR (at $0.1\%$ BER) compared to the jammer-free JL-LMMSE receiver.
\begin{remark}
Reserving time slots for jammer estimation in which the UEs \blue{cannot} transmit
reduces the achievable data~rates.	
\end{remark}

Contrastingly, in \fref{fig:example:fail} the attacking (smart) jammer is aware of the POS receiver's mitigation scheme
and suspends transmission during the time slots that are used to estimate its subspace.
The POS receiver's subspace estimate is thus based entirely on noise and is completely independent
of the jammer's true channel~$\Hj$. Consequently, the mitigation mechanism fails spectacularly, yielding
a bit error-rate identical to the non-mitigating LMMSE receiver.

\section{Joint Jammer Estimation and Mitigation, Channel Estimation, and Data Detection}
The foregoing example has demonstrated the danger of estimating the jammer's subspace
(or other characteristics of~the jammer, such as its spatial covariance) 
based on a certain subset of receive samples when facing a smart jammer.
We therefore propose a method that does not depend on the jammer being active during any specific period.
This independence is achieved by considering the receive signal over an entire coherence interval 
at once and exploiting the fact that the jammer subspace stays fixed within that period, 
regardless of the jammer's activity pattern or transmit sequence. 
Specifically, we first propose a novel optimization problem that combines a tripartite goal of 
(i) mitigating the jammer's interference by locating its subspace $\textit{span}(\Hj)$
and projecting the receive matrix $\bY$ onto the orthogonal complement of~that subspace, 
(ii) estimating the channel matrix~$\bH$, and (iii) recovering the data matrix $\bS_D$.
We then establish the soundness of the proposed optimization problem by proving that, 
under certain sensible conditions, and assuming negligible thermal noise, 
the minimum is unique and corresponds to the desired solution; in particular, 
solving the problem recovers the data matrix~$\bS_D$.   
Finally, we develop efficient iterative algorithms that approximately solve the proposed optimization problem.

\subsection{The Optimization Problem}

\blue{To motivate our optimization problem, we will make the simplifying assumption
that the thermal noise is so low as to be negligible: \mbox{$\bN\approx\mathbf{0}$}.
The reason for making the low-noise assumption is not that it is ultimately necessary for our method to work
(indeed, our numerical results in \fref{sec:results}~show our method to work well also when the noise 
is not negligible), but since the absence of noise helps to understand the problem, and since 
any sensible jammer mitigation method should also work in the absence of additional thermal noise.
}

We start our derivation by considering the maximum-likelihood \blue{(ML)} problem for joint channel estimation and data detection 
\blue{(JED), assuming---in a first step---no jamming activity (i.e., assuming $\bmw=\mathbf{0}$). 
In that case, the ML JED problem is~\cite{vikalo2006efficient}} 
\begin{align}
	 \big\{\hat\bH, \hat\bS_D\big\}
	&= \argmin_{\substack{\hspace{1.3mm}\tilde\bH\in\opC^{B\times U}\\ \tilde\bS_D\in\setS^{U\times D}}}\!
	\big\|\bY - \tilde\bH \tilde\bS \big\|^2_F, \label{eq:ml_jed}
\end{align}
where we define $\tilde\bS \triangleq [\bS_T,\tilde\bS_D]$ for brevity and leave the dependence on
$\tilde\bS_D$ implicit. This objective already integrates the goals of estimating the channel matrix
and detecting the data symbols: If the noise $\bN$ is small enough to be negligible, 
the problem is minimized by the true channel and data matrices,
\begin{align}
	\|\bY - \bH \bS \|^2_F \blue{=\|\bN\|_F^2} \approx 0, \label{eq:no_jamming}
\end{align}
where the pilot matrix $\bS_T$ ensures uniqueness.\footnote{If the noise $\bN$ 
is not strictly equal to zero, then the channel estimate $\hat\bH$ for which \eqref{eq:ml_jed} is minimized
does not coincide \emph{exactly} 
with the true channel matrix~$\bH$. But thanks to the discrete search space, the minimizing data 
estimate $\hat\bS_D$ still coincides exactly with the true data matrix $\bS_D$ if $\bN$ is small enough.}
\blue{Let us now consider how \eqref{eq:no_jamming} is affected by the presence of significant jamming activity: 
$\|\bmw\|_2^2\gg0$. In that case, the jammer will cause a residual}  
\blue{
\begin{align}
	\|\bY - \bH\bS\|^2_F &= \|\Hj\tp{\bsj} + \bN\|^2_F \\
	&\approx \|\Hj\tp{\bsj}\|^2_F \gg 0 \label{eq:jed_residual}
\end{align}
\!\!\!}
when plugging the true channel and data matrices into \fref{eq:ml_jed}. 
\blue{The step in \eqref{eq:jed_residual} follows because we assumed $\|\bmw\|_2^2\gg0$ and $\bN\approx\mathbf{0}$.}
\blue{Considering \eqref{eq:jed_residual}, there might now be a tuple
$\{\tilde\bH,\tilde\bS_D\}$ with $\tilde\bS_D\neq\bS_D$ such that 
$\|\bY - \tilde\bH \tilde\bS\|^2_F < \|\bY - \bH\bS\|^2_F$.}

Note, however, that the residual $\Hj\tp{\bsj}$ in \eqref{eq:jed_residual} is a rank-one matrix whose
columns are all \blue{contained} in $\textit{span}(\Hj)$, regardless of the jamming signal~$\bsj$.
Consider therefore what happens when we take the~matrix\footnote{The dependence
of \blue{$\tilde\bP(\tilde\bmp)$} on $\tilde\bmp$ is left implicit here and throughout the~paper.}
\begin{align}
\tilde\bP\triangleq \blue{\bI-\tilde\bmp\pinv{\tilde\bmp}=~} \bI-\tilde\bmp\herm{\tilde\bmp},~\tilde\bmp\in \mathbb{S}_1^B,
\end{align} 
which projects a signal 
onto the orthogonal complement of some arbitrary one-dimensional subspace $\textit{span}(\tilde\bmp)$\blue{\cite[Sec. 2.6.1]{GV96}},
and then apply that projection to the objective of \eqref{eq:ml_jed}: 
\begin{align}
	\|\tilde\bP(\bY - \tilde\bH\tilde\bS)\|^2_F. \label{eq:ml_p_jed}
\end{align}
If we now plug the true channel and data matrices into \fref{eq:ml_p_jed} (still assuming negligibility of the noise $\bN$), then we obtain
\begin{align}
	\|\tilde\bP(\bY - \bH\bS)\|^2_F 
	&= \|\tilde\bP\Hj\tp{\bsj} + \tilde\bP\bN\|^2_F \\
	&\approx  \|\tilde\bP\Hj\tp{\bsj}\|^2_F \geq 0, 
\end{align}
with equality if and only if $\tilde\bmp$ is collinear with $\Hj$. 
In other words, the unit vector $\tilde\bmp$ which in combination with the true channel and data matrices minimizes \eqref{eq:ml_p_jed}
is collinear with the jammer's channel. 
\blue{In this case, $\tilde\bP = \bI_B - (\Hj/\|\Hj\|_2)(\herm{\Hj}/\|\Hj\|_2) = \bI_B - \Hj\pinv{\Hj}$ 
coincides with the matrix~$\bP_\Hj$ from~\eqref{eq:pos} which projects onto the orthogonal complement of the jammer's subspace.}

Thus, if the noise $\bN$ is negligible, and if
(i)~$\tilde\bP$ is the projection onto the orthogonal complement of $\textit{span}(\Hj)$,
(ii)~$\tilde\bH$ is the true channel matrix,
and (iii) $\tilde\bS$ contains the true data matrix, 
then the tuple $\{\tilde\bmp,\tilde\bH,\tilde\bS\}$ minimizes \eqref{eq:ml_p_jed}.
These~are, of course, exactly the goals which we want to attain.
We thus formulate our joint jammer estimation and mitigation, channel estimation, and data detection problem as follows:
\begin{align}
	 \big\{\hat\bmp, \hat\bH_\bP, \hat\bS_D\big\}
	&= \argmin_{\substack{\tilde\bmp\in \mathbb{S}_1^B\hspace{1.4mm}\\ \hspace{1.3mm}\tilde\bH_\bP\in\opC^{B\times U}\\ \tilde\bS_D\in\setS^{U\times D}}}\!
	\big\|\tilde\bP\bY - \tilde\bH_\bP \tilde\bS \big\|^2_F.\!
	\label{eq:obj1}
\end{align}
Note that, compared to \eqref{eq:ml_p_jed}, we have absorbed the projection matrix $\tilde\bP$ directly into 
the unknown channel matrix $\tilde\bH_\bP$, which replaces the product $\tilde\bP\tilde\bH$ in \eqref{eq:ml_p_jed}. 
\blue{Otherwise, the columns~of~$\tilde\bH$ would 
not be fully determined, since---because of the projection~$\tilde\bP$---the 
magnitude of their components in the direction of 
$\tilde\bmp\approx\Hj$ would be undetermined: 
The objective in \eqref{eq:ml_p_jed} would be unable to distinguish between two
different channel estimates~$\tilde\bH$ and $\tilde\bH'$ when 
$\tilde\bH - \tilde\bH' = \bmj\tp{\tilde\bmw}$ for some $\tilde\bmw \in \opC^U$.}

\subsection{Theory}\label{sec:theory}
We have derived the optimization problem \eqref{eq:obj1} based on intuitive but non-rigorous arguments.
Thus, we will now support the soundness of \eqref{eq:obj1} by proving that, under certain sensible conditions, 
and assuming that the noise is negligible, its solution is unique 
and guaranteed to recover the true data matrix.
\blue{The assumption of negligible noise can be understood as a limiting case of a high SNR scenario.}

We make the following assumptions: 
The channel matrix $\bH$ has full column rank $U$, 
 the jammer channel $\Hj$ is not included in the \blue{column space} of $\bH$, 
and the pilot matrix $\bS_T$ has full row rank $U$. 
In addition, we define a concept which may seem cryptic at first, but which will be clarified later.

\begin{defi} \label{def:eclipse}
\blue{The jammer is \emph{eclipsed} in a given coherence interval if there exists a matrix $\tilde{\bS}_D\in\setS^{U\times D}$, \mbox{$\tilde{\bS}_D\neq \bS_D$,}
such that the matrix \textnormal{$\boldsymbol{\Sigma}\!\triangleq\![\bS_D - \tilde{\bS}_D; \tp{\bsj_D} - \tp{\bsj_T}\pinv{\bS_T}\tilde{\bS}_D]$}
has rank~one.}
\end{defi}
We can now state our result; the proof is in \fref{app:proof1}.
\begin{thm} \label{thm:maed}
In the absence of noise, $\bN=\mathbf{0}$, and if the jammer is not eclipsed, then the problem in \eqref{eq:obj1} has the unique solution $\{\hat\bmp, \hat\bH_\bP, \hat\bS_D\}=\{\bmp, \bP\bH, \bS_D\}$
(In fact, $\hat{\bmp}$ is unique only up to an immaterial phase shift, $\hat{\bmp}=\alpha\bmp, |\alpha|=1$.)
\end{thm}

In other words, as long as the jammer is not ecplised, the problem in \eqref{eq:obj1} is uniquely minimized by the true jammer subspace, projected channel matrix, and data matrix. 
We now shed light on the notion of eclipsedness. 
\blue{We will also show in \fref{thm:maed2}
that---if randomized pilots are used---the jammer is typically not eclipsed in almost all cases. In fact, the jammer is typically not eclipsed even when deterministic pilots are used.}

}

\blue{Eclipsing describes the existence of a certain relationship between the signal and the jamming 
subspace that creates an ambiguity when trying to resolve between the two.} 
In essence, the jammer is eclipsed if its jamming signal $\bmw$ is such that multiple possible ``explanations''
of the receive signal $\bY$ exist which are consistent with the pilot matrix $\bS_T$ 
and under some of which the jammer is not recognized as the jammer; cf. the discussion of \eqref{eq:eclipsing_equation}
in \fref{app:proof1}.
This is best explained by considering two emblematic cases of an eclipsed jammer:
\subsubsection{An inactive jammer (or no jammer)} 
Clearly, if $\bsj=\mathbf{0}$, then 
\blue{the last row of $\boldsymbol{\Sigma}$ is zero for all $\tilde{\bS}_D$, 
including those that differ from $\bS_D$ only in a single row or column, so that eclipsing occurs.}
In this case, there is a mismatch between the jammerless actual wireless transmission and the jammed model in \eqref{eq:io}. 
Since there is no jammer subspace to identify, the choice of the projection~$\tilde\bP$ is undetermined, so that 
\fref{thm:maed} no longer~applies.
\blue{Interestingly, this degenerate case implies that our jammer mitigation method may in fact \emph{require}
the presence of jamming to operate at full effectiveness. In this regard, see 
also the jammerless experiment in \fref{sec:results:eclipsed}.
}

\subsubsection{The jammer transmits a valid pilot sequence}
If the jammer \blue{knows the pilots $\bS_T$ and} transmits the $k$th UE's pilot sequence in the training phase and constellation symbols
in the data phase, then there are no formal grounds for the receiver to distinguish between the jammer and the $k$th UE. 
It can readily be shown that, besides the desired solution 
$\{\hat\bmp, \hat\bH_\bP, \hat\bS_D\}=\{\bmp, \bP\bH, \bS_D\}$, there exists then another solution to~\eqref{eq:obj1}
which identifies the $k$th UE as the jammer, nulls that UE by setting $\hat\bmp = \bmh_k/\|\bmh_k\|$, 
and instead identifies the jammer as the $k$th UE by estimating
\begin{align}
	\hat\bH_\bP &= \hat\bP [ \bmh_1, \dots, \bmh_{k-1}, \Hj, \bmh_{k+1}, \dots, \bmh_U ], \\
	\hat\bS_D &= \tp{[ \tp{\bms_{D,1}}, \dots, \tp{\bms_{D,k-1}}, \tp{\bmw_D}, \tp{\bms_{D,k+1}}, \dots, \tp{\bms_{D,U}} ]},
\end{align}
where $\bms_{D,u}$ is the $u$th row of $\bS_D$.
\blue{This is a case of eclisping, since for $\hat\bS_D = \tilde\bS_D$,
all rows of $\boldsymbol{\Sigma}$ except the $u$th row are zero, so that $\boldsymbol{\Sigma}$ has rank one.}

\blue{Besides these two paradigmatic cases, eclipsing may also happen ``accidentally'' in cases 
where, for some $\tilde\bS_D$, the symbol error matrix $\bS_D - \tilde\bS_D$ has rank one
and its rows are, by coincidence, all collinear with $\tp{\bsj_D} - \tp{\bsj_T}\pinv{\bS_T}\tilde{\bS}_D$.}

However, we will now show that if the jammer does not know the pilot sequences, e.g., because
they are drawn at random by the BS and secretly communicated to the UEs, then an active jammer (where $\bsj\neq \mathbf{0}$)
is typically not eclipsed. 
\blue{Thus, to obtain the best possible resilience against smart jammers, randomized pilots should
be used.}
To show this, we consider a case in which the pilot matrix $\bS_T$ is square; the proof is relegated to \fref{app:proof2}.

\begin{thm} \label{thm:maed2}
	\blue{If the pilot matrix $\bS_T$ is drawn uniformly over the set of $U\times U$ unitary matrices and if 
	$\bsj_T\neq\mathbf{0}$ and $\bsj_D\neq\mathbf{0}$ are
	independent of $\bS_T$,	then the probability that the jammer eclipses is bounded from above 
	by $|\setS|^{3U}|\setS|^{-(U-3)D}$, i.e., the probability of eclipsing decreases
	exponentially in the number~$D$ of data time slots processed simultaneously.}
\end{thm}

\begin{exmpl}
	\blue{If the assumptions of \fref{thm:maed2} are satisfied, and if
	$\setS$ is $16$-QAM, $U=32$ and $D=128$, as in most of our experiments in \fref{sec:results}, 
	then the probability of eclipsing is at most $16^{-3616}\approx 10^{-4338}$.
	Even if $\setS$ is QPSK, $U=4$, and one processes only $D=20$ 
	data slots simultaneously, the probability of eclipsing is bounded by $1.6\times10^{-5}$.
	}
\end{exmpl}

\begin{remark} \label{rem:rare}
	\blue{It is by no means necessary to use random pilots to avoid eclipsing. 
	Nor is it necessary that both $\bsj_T$ \emph{and} $\bsj_D$ are distinct from zero. 
	Another sufficient condition for the jammer to be eclipsed only with zero probability
	is, e.g., if $\bsj$ has at least two independent marginals with continuous distribution.
	The main point is that, unless the jammer choses its input sequence as some (partially randomized) 
	function of the pilot matrix $\bS_T$, eclipsing is the rare exception, not the norm. 
	In this regard, see also the simulation results in \fref{sec:results}.}
\end{remark}

\begin{remark} 
The fact that \blue{reliable} communication in the presence of jamming can be assured if 
the BS and UEs~share~a common secret that enables them to use a randomized~communication scheme, but 
not otherwise, is reminiscent of~information-theoretic results which prove a similar 
dichotomy on a more fundamental level. See \cite[Sec. V]{lapidoth1998reliable} and references therein.
\end{remark}

\section{Forward-Backward Splitting with a Box Prior}

We now provide the first of two algorithms for approximately solving the
joint jammer estimation and mitigation, channel estimation, and data detection problem in \eqref{eq:obj1}. 
Note first of all that the objective is quadratic in $\tilde\bH_\bP$, 
so we can derive the optimal value of $\tilde\bH_\bP$ as a function of $\tilde\bP$ and $\tilde\bS$ as
\begin{align}
	\hat\bH_\bP = \tilde\bP\bY\pinv{\tilde\bS}, 
\end{align}
where $\pinv{\tilde\bS}=\herm{\tilde\bS}\inv{(\tilde\bS\herm{\tilde\bS})}$. 
Substituting $\hat\bH_\bP$ back into \eqref{eq:obj1} yields 
an optimization problem which only depends on $\tilde\bmp$ and $\tilde\bS_D$:
\begin{align}
	\big\{\hat\bmp, \hat\bS_D\big\} = 
	\argmin_{\substack{\tilde\bmp\in \mathbb{S}_1^B\hspace{1.4mm}\\ \tilde\bS_D\in\setS^{D\times U}}}
	\big\|\tilde\bP\bY(\bI_K - \pinv{\tilde\bS}\tilde\bS)\big\|^2_F. \label{eq:obj3}
\end{align}
Solving \eqref{eq:obj3} remains difficult due to its combinatorial nature, so we resort to solving it approximately. 
First, we relax the constraint set $\setS$ to its convex hull $\setC\triangleq\textit{conv}(\setS)$ as in \cite{castaneda2018vlsi}.
This can be viewed as replacing the probability mass function over the constellation
$\setS$, which represents the true symbol prior, with a box prior that is uniform 
over $\setC$ and zero elsewhere \cite{jeon2021mismatched}. 
We then approximately solve this~relaxed problem formulation in an iterative fashion by alternating between 
a forward-backward splitting descent step in $\tilde\bS$ and a minimization step in $\tilde\bP$.

\subsection{Forward-Backward Splitting Step in $\tilde\bS$} \label{sec:fbs}
Forward-backward splitting (FBS) \cite{goldstein16a}, also called proximal gradient descent\blue{\cite{parikh13a}},
is an iterative method for solving convex optimization problems of the form
\begin{align}
	\argmin_{\tilde\bms}\, f(\tilde\bms) + g(\tilde\bms), \label{eq:fbs1}
\end{align}
where $f$ is convex and differentiable, and $g$ is convex but not necessarily
differentiable, smooth, or bounded. Starting from an initialization vector $\tilde\bms^{(0)}$, 
FBS solves the problem in~\eqref{eq:fbs1} iteratively by computing 
\begin{align}
	\tilde\bms^{(t+1)} = \proxg\big(\tilde\bms^{(t)} - \tau^{(t)}\nabla f(\tilde\bms^{(t)}); \tau^{(t)}\big). \label{eq:fbs2}
\end{align}
Here, $\tau^{(t)}$ is the stepsize at iteration $t$, $\nabla f(\tilde\bms)$ is the gradient~of $f(\tilde\bms)$,  
and $\proxg$ is the proximal operator of $g$, defined as \cite{parikh13a}
\begin{align}
	\proxg(\bmx; \tau) = \argmin_{\tilde\bmx} \tau g(\tilde\bmx) + \frac12 \|\bmx - \tilde\bmx\|_2^2.
\end{align}
For a suitable sequence of stepsizes $\{\tau^{(t)}\}$, FBS solves convex optimization problems exactly.
FBS can also be used to approximately solve non-convex
problems, although there are typically no guarantees for optimality or even convergence~\cite{goldstein16a}.
For the optimization problem in \fref{eq:obj3}, we define $f$ and $g$ as 
\begin{align}
	f(\tilde\bS) &= \big\|\tilde\bP\bY(\bI_K - \pinv{\tilde\bS}\tilde\bS)\big\|^2_F
\end{align}
and
\begin{align}
	g(\tilde\bS) &= \begin{cases}
		0 &\text{if }\,\tilde\bS_{[1:T]}=\bS_T \text{ and } \tilde\bS_{[T+1:K]}\in\setC^{U\times D}
		\!\!\!\\
		\infty &\text{else}.
	\end{cases}
\end{align}
The gradient of $f$ in $\tilde\bS$ is given by 
\begin{align}
	\nabla f(\tilde\bS) = -\herm{(\bY\pinv{\tilde\bS})}\tilde\bP\bY(\bI_K - \pinv{\tilde\bS}\tilde\bS), \label{eq:gradient}
\end{align}
and the proximal operator for $g$ is simply the orthogonal projection onto $\setC$, which  
acts entrywise on $\tilde\bS$~as
\begin{align}
	[\proxg(\tilde\bS; \tau)]_{u,k} = \begin{cases}
		[\bS_T]_{u,k} &\text{ if } k\in[1:T] \\
		\text{proj}_\setC([\tilde\bS]_{u,k}) &\text{ else,}
	\end{cases} \label{eq:proxg} 
\end{align}
where the function $\text{proj}_\setC$ is given as 
\begin{align}
	\text{proj}_\setC(x) =\, & \min\{\max\{\Re(x),-\lambda\},\lambda\} \nonumber\\
	&+ i\min\{\max\{\Im(x),-\lambda\},\lambda\},
\end{align}
\blue{where $\lambda$ denotes the edge of the constellation's convex hull~$\setC$, see \fref{fig:constellations}. The
value of $\lambda$ is determined by the fact that the transmit constellations are scaled 
to unit average symbol energy (cf. \fref{sec:setup}). Specifically, we have
$\lambda=\sqrt{\sfrac{1}{2}}$ for a QPSK constellation and $\lambda=\sqrt{\sfrac{9}{10}}$ for a 16-QAM constellation.}
To select the per-iteration stepsizes~$\{\tau^{(t)}\}$, we use the Barzilai-Borwein method 
\cite{barzilai1988two}.

\subsection{Minimization Step in $\tilde\bP$}
After each FBS step in $\tilde\bS$, we minimize \eqref{eq:obj3}
with respect to the vector~$\tilde\bmp$. Defining the residual matrix
$\tilde\bE\triangleq \bY(\bI_K - \pinv{\tilde{\bS}}\tilde{\bS})$
and performing standard algebraic manipulations yields
\begin{align}
	\hat\bmp &= \argmin_{\tilde\bmp\in \mathbb{S}_1^B} \big\|\tilde\bP \tilde\bE \big\|^2_F	\\
	&= \argmax_{\tilde\bmp\in \mathbb{S}_1^B} \, \herm{\tilde\bmp} \tilde\bE \herm{\tilde\bE} \tilde\bmp. \label{eq:rayleigh}
\end{align}
It follows that the vector $\hat\bmp$ minimizing \eqref{eq:obj3} for a fixed~$\tilde\bS$ is the unit vector that maximizes the 
Rayleigh quotient of $\tilde\bE \herm{\tilde\bE}$. 
The~solution is the \blue{unit-$2$-norm} eigenvector 
$\bmv_1(\tilde\bE \herm{\tilde\bE})$ associated with the largest eigenvalue of $\tilde\bE \herm{\tilde\bE}$,
\begin{align}
	\hat\bmp=\bmv_1(\tilde\bE \herm{\tilde\bE}).
\end{align}
Calculating this eigenvector in every iteration of our algorithm would be computationally expensive, 
so we approximate it using a single power iteration \cite[Sec.\,8.2.1]{GV96}, i.e., 
we estimate 
\begin{align}
	\hat\bmp^{(t+1)} = \frac{\tilde\bE^{(t+1)} \herm{(\tilde\bE^{(t+1)})}\hat\bmp^{(t)}}{\|\tilde\bE^{(t+1)} \herm{(\tilde\bE^{(t+1)})}\hat\bmp^{(t)}\|_2},
\end{align}
where the power method is initialized with the subspace estimate $\hat\bmp^{(t)}$ from the previous algorithm iteration.

\subsection{Preprocessing}
If the algorithm starts directly with a gradient descent step in the direction of \eqref{eq:gradient}, 
one runs the risk of advancing significantly into the wrong direction---especially if the jammer is extremely strong,
since a strong jammer will also lead to a large gradient amplitude. Empirically, we observe that such a large initial digression 
can be problematic (if, e.g., the jammer is $\geq\!50$\,dB stronger than the average UE).
It might therefore be tempting to start the algorithm directly with a projection step: 
If one initializes $\tilde\bS^{(0)}=\mathbf{0}_{U\times D}$, then $\tilde\bE^{(0)}=\bY$, so that the algorithm starts by 
nulling the dimension of $\bY$ which contains the most energy. In the presence of a strong jammer, this is a sensible strategy
since this dimension then corresponds to the jammer subspace. However, if the received jamming energy is 
small compared to the energy received from the UEs (e.g., because the jammer does not transmit at all during a given coherence interval), 
then such a projection would inadvertently null the strongest user. 
To thread the needle between these two cases---largely removing a strong jammer before the first gradient step, 
but not removing any legitimate UEs when a strong jammer is absent---we propose to start with a \emph{regularized} projection step:
The algorithm starts by a projection onto the orthogonal complement of the eigenvector of the largest eigenvalue
of 
\begin{align}
	\bY\herm{\bY} + \mathbf{\Gamma}, \label{eq:regularizer}
\end{align}
where $\mathbf{\Gamma}\in\opC^{B\times B}$ is a constant regularization matrix. The basic idea is that this regularization matrix is
still overshadowed by very strong jammers, so that these are largely nulled within the preprocessing, 
while, in the presence of only a weak jammer (or no jammer), the regularization matrix has a sufficiently diverting impact 
on the eigenvectors to prevent the nulling of a legitimate UE. 
There are countless ways of choosing such a regularization matrix. (Note, however, that $\mathbf{\Gamma}$ should 
not be a multiple of the identity matrix $\bI_B$, which does not affect the eigenvectors of \eqref{eq:regularizer}.) For simplicity, we set $\mathbf{\Gamma}$ to the all-zero matrix, except for 
the top left entry, which is set to $0.1BUK$. 

\subsection{Algorithm Complexity} 
We now have all the ingredients for MAED, which is summarized in \fref{alg:maed}. Its only input is the receive matrix~$\bY$, 
as it does not even require an estimate of the thermal noise variance $\No$.
MAED is initialized with $\tilde\bS^{(0)}=[\bS_T, \mathbf{0}_{U\times D}]$ and $\tau^{(0)}=\tau=0.1$, 
and runs for a fixed number of $t_{\max}$~iterations.

The complexity of MAED is dominated by the eigenvector calculation in the preprocessing step
(which could be reduced \blue{by using the power method approximation}) as well 
as the gradient computation in line 5 of \fref{alg:maed},  
which has a complexity of $O(3BUK+2U^2K+U^3)$.
The overall complexity of MAED is therefore \mbox{$O(t_{\max}(3BUK+2U^2K+U^3))$.}
Note, however, that MAED detects $D$ data vectors at once.
Thus, the computational complexity per detected symbol is only \mbox{$O(t_{\max}(3BK+2KU+U^2)/D)$.}

\section{Soft-Output Estimates with Deep Unfolding}
MAED, which corresponds to the algorithm proposed in~\cite{marti2022smart} \blue{(adding the new preprocessing step)},
already attains the goal of mitigating smart jammers, see \fref{sec:results}. 
However, its detection performance can be suboptimal, especially when higher-order transmit constellations 
such as 16-QAM are used. 
The culprit is the box prior of MAED, which does not fully exploit the 
discrete nature of the transmit constellation. In particular, the box prior is uninformative about 
the  constellation symbols in the interior \blue{of $\setC$.}
To improve detection performance,
we now provide a second algorithm for approximately solving the problem 
in \eqref{eq:obj1}. This second algorithm builds on MAED but replaces the proximal operator in \eqref{eq:proxg}, which enforces
MAED's box prior, by an approximate posterior mean estimator (PME) based on the discrete symbol prior as in \cite{song2021soft}.  
Since the PME also enables meaningful soft-output estimates of the bits that underlie the transmitted data symbols, we refer to 
this second algorithm as soft-output MAED (SO-MAED).

\begin{algorithm}[tp]
\caption{MAED}
\label{alg:maed}
\begin{algorithmic}[1]
\setstretch{1.0}
\State {\bfseries input:} $\bY$
\State \text{initialize} $\tilde\bS^{(0)} \!=\! \left[\bS_T, \mathbf{0}_{U\!\times\! D} \right]\!, 
\tilde\bmp^{(0)} \!=\! \bmv_1(\bY \herm{\bY} \!+ \mathbf{\Gamma} ), \tau^{(0)} \!= \tau$
\State $\tilde\bP^{(0)} = \bI_B - \tilde\bmp^{(0)}\tilde\bmp^{(0)}{}^\text{H}$
\For{$t=0$ {\bfseries to} $t_{\max}-1$}
	\State $\nabla f(\tilde\bS^{(t)}) = -\herm{\big(\bY\tilde\bS^{(t)}{}^\dagger\big)} \tilde\bP^{(t)}\bY(\bI_K - \tilde\bS^{(t)}{}^\dagger \tilde\bS^{(t)})$
	\State $\tilde\bS^{(t+1)} = \proxg\big(\tilde\bS^{(t)} - \tau^{(t)}\nabla f(\tilde\bS^{(t)})\big)$ 
	\State $\tilde\bE^{(t+1)} = \bY(\bI_K - \tilde\bS^{(t+1)}{}^\dagger \tilde\bS^{(t+1)})$
	\State $\tilde\bmp^{(t+1)} = \tilde\bE^{(t+1)} \tilde{\bE}^{(t+1)}{}^\text{H}\, \tilde\bmp^{(t)}/\|\tilde\bE^{(t+1)} \tilde{\bE}^{(t+1)}{}^\text{H}\, \tilde\bmp^{(t)}\|_2$
	\State $\tilde\bP^{(t+1)} = \bI_B - \tilde\bmp^{(t+1)}\tilde\bmp^{(t+1)}{}^\text{H}$
	\State compute $\tau^{(t+1)}$ by following \cite[Sec.\,4.1]{goldstein16a}
	\EndFor
	\State \textbf{output:} $\tilde\bS^{(t_{\max})}_{[T+1:K]}$
\end{algorithmic}
\end{algorithm}

\subsection{Approximate Posterior-Mean Estimation}

To replace the proximal operator following the gradient descent step in \eqref{eq:fbs2}
with a more appropriate data symbol estimator which takes into account the discrete constellation~$\setS$,
we model the per-iteration outputs of the gradient descent step
\begin{align}
	\tilde\bX^{(t)} = \tilde\bS^{(t)} - \tau^{(t)}\nabla f(\tilde\bS^{(t)})
\end{align} 
as 
\begin{align}
	\tilde\bX^{(t)} &= \bS + \bZ^{(t)} = \big[ \bS_T, \bS_D \big] + \big[ \bZ_T^{(t)}, \bZ_D^{(t)} \big], \label{eq:symbol_noise}
\end{align}
i.e., as the true transmit symbol matrix $\bS$ corrupted by an additive error~$\bZ^{(t)}$.
If the distribution of $\bZ^{(t)}$ were known, 
one could compute the posterior mean $\mathbb{E}[\bS\,|\,\tilde\bX^{(t)}]$
and use it as a constellation-aware replacement of the proximal step \eqref{eq:proxg}, 
\begin{align}
	\tilde\bS^{(t+1)} = \mathbb{E}\big[\bS\,|\,\tilde\bX^{(t)}\big].
\end{align}
Unfortunately, the distribution of $\bZ^{(t)}$ is unknown in practice.
\blue{Calculating the conditional mean of the submatrix $\bS_T$ is nonetheless trivial, 
since $\bS_T$ is deterministically known at the receiver,~so that $\mathbb{E}[\bS\,|\,\tilde\bX^{(t)}]=\bS_T$.}
To estimate the mean of the submatrix $\bS_D$, we assume that
the entries of $\bZ_D^{(t)}$ are distributed independently~of $\bS$ as i.i.d.\ circularly-symmetric 
complex Gaussians with variance~$\nu^{(t)}$, 
\begin{align}
	\big[\bZ^{(t)}\big]_{u,k}\sim\setC\setN(0,\nu^{(t)}).
\end{align}
The variances $\{\nu^{(t)}\}_{t=0}^{t_{\max}-1}$ are treated as algorithm parameters
and will be optimized using deep unfolding as detailed below. 

Based on this idealized model, we \blue{use} a three-step procedure as in \cite{song2021soft} \blue{to compute
symbol estimates:} First, we use \fref{eq:symbol_noise} to compute log-likelihood ratios (LLRs)
for every transmitted bit. We then convert these LLRs to the probabilities of the respective bits being $1$. 
This step also provides the aforementioned soft-output estimates. 
Finally, we convert the bit probabilities back to symbol estimates by calculating the symbol mean. 
 
\begin{figure}[tp]
\centering

\subfigure[QPSK\hspace{1cm}]{
\includegraphics[height=3.4cm]{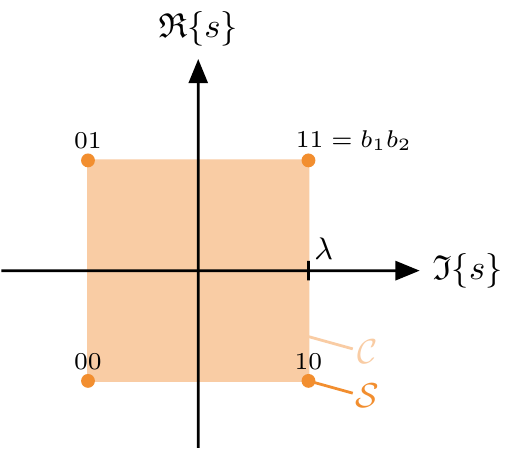}
\label{fig:constellations:qpsk}
}
\subfigure[16-QAM\hspace{1cm}]{
\includegraphics[height=3.4cm]{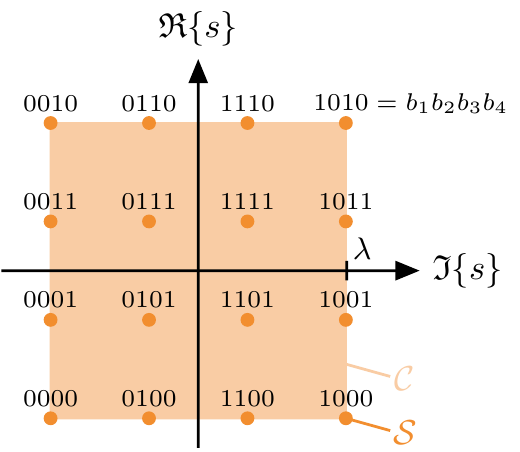}
\label{fig:constellations:16qam}
}
\caption{Transmit constellations $\setS$ (including the used Gray mapping) and their convex hulls $\setC$. 
$\lambda=\sqrt{\sfrac{1}{2}}$ for QPSK and $\lambda=\sqrt{\sfrac{9}{10}}$ for 16-QAM.
}
\vspace{-2mm}
\label{fig:constellations}
\end{figure}

\blue{From \eqref{eq:symbol_noise}, the} LLRs can be computed following \cite{collings2004low, jeon2021mismatched}~as
\begin{align}
	L_{i,u,k}^{(t)} = \frac{\ell\big(\tilde X_{u,k}^{(t)}\big)\!}{\nu^{(t)}}, 
	~ i\!\in\![1\!:\!\log_2\!|\setS|],~ u\!\in\![1\!:\!U],~ k\!\in\![T\!+\!1\!:\!K],
		 \label{eq:llrs}
\end{align}
where $\ell$ is specified in Table~\ref{tab:llr} (cf.~\fref{fig:constellations}).
The LLR values are exact for QPSK and use the max-log 
approximation for 16-QAM~\cite{jeon2021mismatched}.
The LLRs can then be converted to probabilities~via
\begin{align}
	p_{i,u,k}^{(t)} = \frac12\left( 1 + \tanh{\left(\frac{L_{i,u,k}^{(t)}}{2}\right)} \right).
	\label{eq:bit_probabs}
\end{align}
Finally, the probabilities of \eqref{eq:bit_probabs} can be used to compute symbol estimates according to Table~\ref{tab:symbol_estimates}.

To summarize, SO-MAED replaces MAED's proximal operator in \eqref{eq:proxg}
with the symbol estimator that consists of \eqref{eq:llrs}, \eqref{eq:bit_probabs}, and Table~\ref{tab:symbol_estimates}.
We refer to this symbol estimation as posterior-mean approximation (PMA) and denote it as 
\begin{align}
	\tilde\bS^{(t+1)} = \pma(\tilde\bX^{(t)},\nu^{(t)}),		
\end{align}
where the subscript $\setS$ makes explicit the dependence of the PMA on the symbol constellation.
Since the PMA involves only scalar computations, its complexity is negligible compared to the
matrix-vector and matrix-matrix operations of SO-MAED. The complexity order of SO-MAED is 
therefore identical to that of MAED, namely \mbox{$O(t_{\max}(3BUK+2U^2K+U^3))$.}

\subsection{Deep Unfolding of SO-MAED}

The procedure outlined in the previous subsection requires the variances $\{\nu^{(t)}\}_{t=0}^{t_{\max}-1}$ 
of the per-iteration estimation errors~$\bZ^{(t)}$, which are generally unknown. We treat these variances 
as parameters of SO-MAED and optimize them using deep unfolding \cite{song2021soft, hershey2014deep, balatsoukas2019deep, goutay2020deep, monga2021algorithm}.
Deep unfolding is an emerging paradigm \blue{in} which iterative algorithms are unfolded into artificial neural networks
with one layer per iteration, so that the algorithm parameters can be
regarded as trainable weights of that network. These weights are then learned from training data with  
standard deep learning tools~\cite{Baydin2018automatic, tensorflow2015whitepaper}.

\begin{figure}[tp]
\centering
\includegraphics[width=\columnwidth]{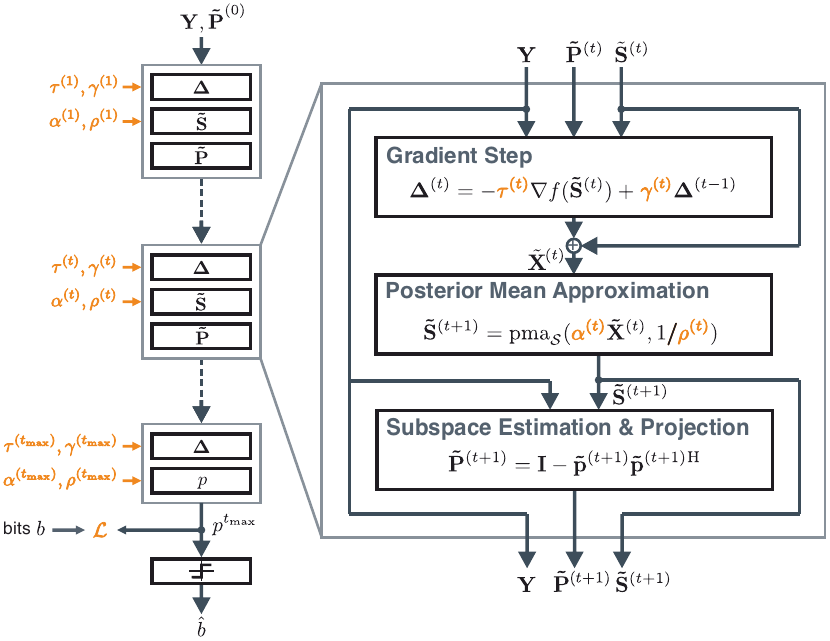}
\vspace{-5mm}
\caption{\blue{A graphical illustration of the neural network which implements the SO-MAED algorithm.
The trainable parameters 
are depicted in orange. Also in orange is the loss function $\setL$ (cf. \eqref{eq:loss}) which is used for training
and which has as inputs the ground-truth transmitted information bits $b$ of the training set $\setD$, 
as well as their respective probabilistic estimates $p^{t_{\max}}$.}
}
\label{fig:so_maed_network}
\vspace{-2mm}
\end{figure}

To improve the stability of learning, we use the error precisions $\{\rho^{(t)}\}_{t=0}^{t_{\max}-1}$ 
instead of the variances $\{\nu^{(t)}\}_{t=0}^{t_{\max}-1}$
as parameters of the unfolded network, with $\rho^{(t)}=1/\nu^{(t)}$.
In addition, we also regard the gradient step sizes $\{\tau^{(t)}\}_{t=0}^{t_{\max}-1}$
as trainable weights (instead of computing them according to the Barzilai-Borwein method). 
Furthermore, we add a momentum term with per-iteration weights \blue{$\{\gamma^{(t)}\}_{t=0}^{t_{\max}-1}$} to our gradient descent procedure.
Finally, inspired by the Bussgang decomposition \cite{bussgang52a, minkoff85a}, we add per-iteration scale factors 
$\{\alpha^{(t)}\}_{t=0}^{t_{\max}-1}$ to the output of \eqref{eq:symbol_noise}, 
with the goal of accommodating uncorrelatedness (if not independence) between $\bZ_D^{(t)}$ and $\bS_D$ in \eqref{eq:symbol_noise}.
The final algorithm is summarized in \fref{alg:so_maed}\blue{, and the 
corresponding unfolded network is visualized in \fref{fig:so_maed_network}.}

\begin{table}[tp!]
\centering
\caption{LLR computation according to \cite[Tbl. 1]{collings2004low}, \cite{jeon2021mismatched}}
\vspace{-5mm}
\setstretch{1.1}
\begin{tabular}[t]{@{}lcl@{}}
\toprule
&\!\!Bit $i$\!\!& $\ell(x)$ \\
\midrule
\multirow{2}{*}{\!QPSK\!}& $1$ & $4\lambda\Re\{x\}$  \\ 
& $2$ & $4\lambda\Im\{x\}$ \vspace{1mm} \\
\multirow{4}{*}{\!16-QAM\!}& $1$ & $\frac{2\lambda}{3} \left( 4\Re\{x\} + \left|\Re\{x\} \!-\! \frac{2\lambda}{3} \right| - \left|\Re\{x\} \!+\! \frac{2\lambda}{3} \right| \right)\!$ \\
& $2$ & $\frac{4\lambda}{3} \left( \frac{2\lambda}{3} - \left|\Re\{x\} \right| \right) $ \\
& $3$ & $\frac{2\lambda}{3} \left( 4\Im\{x\} + \left|\Im\{x\} \!-\! \frac{2\lambda}{3} \right| - \left|\Im\{x\} \!+\!\frac{2\lambda}{3} \right| \right)\! $ \\
& $4$ & $\frac{4\lambda}{3} \left( \frac{2\lambda}{3} - \left|\Im\{x\} \right| \right) $ \\
\bottomrule
\end{tabular}
\vspace{5mm}
\label{tab:llr}
\centering
\caption{Mapping the probabilities in \eqref{eq:bit_probabs} to symbol estimates
\cite{jeon2021mismatched, tomasoni2006low}}
\vspace{-5mm}
\setstretch{1.1}
\begin{tabular}[t]{@{}lcc@{}}
\toprule
& $\Re\{\hat{s}\}$\!& $\Im\{\hat{s}\}$ \\
\midrule
QPSK & $\lambda(2p_{1}-1)$ & $\lambda(2p_{2}-1)$  \\
16-QAM& $\blue{\frac{\lambda}{3}}(2p_1-1)(3-2p_2)$ &  $\blue{\frac{\lambda}{3}}(2p_3-1)(3-2p_4)$ \\
\bottomrule
\end{tabular}
\vspace{-2.5mm}
\label{tab:symbol_estimates} 
\end{table}%

We implement the unfolded algorithm in TensorFlow \cite{tensorflow2015whitepaper}.
\blue{As loss function, we use the empirical binary cross-entropy (BCE) on the training set $\setD$
between the transmitted bits 
and the estimated bit probabilities \eqref{eq:bit_probabs} from the last iteration as the output of our network.
The loss as a function of the weights 
$\boldsymbol{\theta}=\{\tau^{(t)}, \gamma^{(t)}, \alpha^{(t)}, \rho^{(t)}\}_{t=0}^{t_{\max}}$ is therefore} 
\begin{align}
\!\!\!\! \setL(\boldsymbol{\theta})=\!
	\sum_{d=1}^{|\setD|} \beta^{(d)} \!\!\! \sum_{i=1}^{\log_2\!|\setS|} \! \sum_{u=1}^U \sum_{k=T+1}^K \!\!\!
	\text{BCE}\! \left(b_{i,u,k}^{(d)}, \Big(\!p_{i,u,k}^{(t_{\max})}\!\Big)^{\!\!(d)} \right)\!, \!\!\! \label{eq:loss}
\end{align}
where
\begin{align}
	\text{BCE}(b,p) = b \log_2(p) + (1-b)\log_2(1-p),
\end{align}
and where \blue{$\beta^{(d)}$} are the weights given to the different samples in the training set $\setD$ 
\blue{(see below)}.
\blue{The dependence on the right-hand-side of \eqref{eq:loss} on the parameters $\boldsymbol{\theta}$
is through the bit probabilites $p_{i,u,k}^{(t_{\max})}$ which are functions of $\boldsymbol{\theta}$.} 

We only learn a single set of weights per system dimensions $\{U,B,K\}$, which is used for all 
signal-to-noise ratios (SNRs) and, most importantly, 
all jamming attacks (since a receiver does not typically know in advance
which type of a jamming attack it is facing). 
For this reason, we train \blue{using a training set~$\setD$ which contains}
samples from different SNRs and different jamming attacks. 
We also have to avoid overfitting to a specific type of jamming attack.
If our evaluation in \fref{sec:results} would feature only the exact same types of jammers 
that were used for training, this would raise questions about the ability of SO-MAED to generalize to 
jamming attacks which differ from those explicitly included in the training set.
However, the principles underlying the SO-MAED algorithm are essentially invariant with respect to the type of a 
jamming attack. For this reason, we only train on a single type of jammers, namely pilot jammers,\blue{\footnote{\blue{In 
other words, the training set $\setD$ contains only samples in which the jammer is a pilot jammer.}}}
cf. \fref{sec:setup} (which we have empirically recognized to be the most difficult to mitigate), while evaluating the trained algorithm on many other jammer types besides pilot jammers, cf. \fref{sec:results}.
The attacks used for training also comprise different jammer receive strengths,
namely $\{-\infty\,\text{dB}, 0\,\text{dB}, 10\,\text{dB}, 20\,\text{dB}, 40\,\text{dB}, 80\,\text{dB}\}$  
relative to the average UE. 

The sample weights $\beta^{(d)}$ are used to prevent certain training samples (e.g., those at low SNR with strong jammers)
from dominating the learning process by drowning out the loss contribution \blue{of} training samples with inherently lower BCE. 
For this, we fix a baseline performance and select the weight \blue{$\beta^{(d)}$}
of a training sample as the inverse of this sample's BCE loss 
according to the baseline. \blue{The baseline performance} is set by an untrained version of SO-MAED with reasonably initialized weights \blue{$\boldsymbol{\theta}$}
(its performance in general already exceeds that~of~MAED).

For training, we use the Adam optimizer \cite{kingma2014adam} from Keras with default values\blue{\cite{keras_adam}}.
\blue{Training starts with a batch size of one sample, but the batch size is increased (first to five, then 
to ten, and finally to twenty samples) whenever the training loss does not improve for two consecutive
epochs.}

\blue{
For a more extensive discussion on deep learning architectures in communication transceivers, we
refer to \cite{jagannath2021redefining, albreem2021deep}.}

\begin{algorithm}[tp]
\caption{SO-MAED}
\label{alg:so_maed}
\begin{algorithmic}[1]
\setstretch{1.0}
\State {\bfseries input:} $\bY, \{\tau^{(t)}, \alpha^{(t)}, \blue{\gamma^{(t)}}, \rho^{(t)} \}_{t=0}^{t_{\max} -1}$
\State $\tilde\bS^{(0)} \!=\! \left[\bS_T, \mathbf{0}_{U\!\times\! D} \right], 
\tilde\bmp^{(0)} \!=\! \bmv_1(\bY \herm{\bY} \!+ \mathbf{\Gamma} ), \mathbf{\Delta}^{(-1)} = \mathbf{0}$
\State $\tilde\bP^{(0)} = \bI_B - \tilde\bmp^{(0)}\tilde\bmp^{(0)}{}^\text{H}$
\For{$t=0$ {\bfseries to} $t_{\max}-1$}
	\State $\nabla f(\tilde\bS^{(t)}) = -\herm{\big(\bY\tilde\bS^{(t)}{}^\dagger\big)} \tilde\bP^{(t)}\bY(\bI_K - \tilde\bS^{(t)}{}^\dagger \tilde\bS^{(t)})$
	\State $\mathbf{\Delta}^{(t)} = - \tau^{(t)} \nabla f(\tilde\bS^{(t)}) + \blue{\gamma^{(t)}} \mathbf{\Delta}^{(t-1)}$
	\State $\tilde\bX^{(t)} = \tilde\bS^{(t)} + \mathbf{\Delta}^{(t)} $
	\State $\tilde\bS^{(t+1)} = \pma\big( \alpha^{(t)}\tilde\bX^{(t)}, 1/\rho^{(t)} \big)$ 
	\State $\tilde\bE^{(t+1)} = \bY(\bI_K - \tilde\bS^{(t+1)}{}^\dagger \tilde\bS^{(t+1)})$
	\State $\tilde\bmp^{(t+1)} = \tilde\bE^{(t+1)} \tilde{\bE}^{(t+1)}{}^\text{H}\, \tilde\bmp^{(t)}/\|\tilde\bE^{(t+1)} \tilde{\bE}^{(t+1)}{}^\text{H}\, \tilde\bmp^{(t)}\|_2$
	\State $\tilde\bP^{(t+1)} = \bI_B - \tilde\bmp^{(t+1)}\tilde\bmp^{(t+1)}{}^\text{H}$
	\EndFor
	\State \textbf{output:} $\tilde\bS^{(t_{\max})}_{[T+1:K]}$
\end{algorithmic}
\end{algorithm}

\section{Simulation Results} \label{sec:results}
\subsection{Simulation Setup} \label{sec:setup}
We simulate a massive MU-MIMO system with $B=128$~BS antennas, 
$U=32$ single-antenna UEs, and one single-antenna jammer. 
The UEs transmit for $K=160$ time slots, where the first
$T=32$ slots are used for orthogonal pilots $\bS_T$ in the form of a
 $32\times32$ Hadamard matrix with unit symbol energy.
The remaining $D=128$ slots are used to transmit QPSK or 16-QAM payload data.
Unless noted otherwise, the channels are modelled as i.i.d. Rayleigh fading. 
We define the average receive signal-to-noise ratio (SNR) as 
\begin{align}
\textit{SNR} \define \frac{\Ex{\bS}{\|\bH\bS\|_F^2}}{\Ex{\bN}{\|\bN\|_F^2}}.
\end{align}
We consider four different jammer types:  
(J1) barrage jammers that transmit i.i.d.\ jamming symbols during the entire coherence interval,
(J2) pilot jammers that transmit i.i.d.\ jamming symbols during the pilot phase but do not jam the data phase, 
(J3) data jammers that transmit i.i.d.\ jamming symbols during the data phase but do not jam the pilot phase, 
and (J4) sparse jammers that transmit i.i.d.\ jamming symbols during some fraction $\alpha$ of randomly selected bursts of unit length (i.e., one time slot).
The jamming symbols are either circularly symmetric complex Gaussian or drawn uniformly from the transmit constellation 
(i.e., QPSK or 16-QAM).
They are also independent of the UE transmit symbols $\bS$, unless stated otherwise.
We quantify the strength of the jammer's interference relative to the strength of the average UE, either as the ratio 
between total receive~\textit{energy}
\begin{align}
\rE \define \frac{\Ex{\bsj}{\|\Hj\bsj\|_2^2}}{\frac1U\Ex{\bS}{\|\bH\bS\|_F^2}},
\end{align}
or as the ratio between receive \textit{power during those phases that the jammer is jamming}
\vspace{-1mm}
\begin{align}
	\rP \triangleq \frac{\rE}{\gamma},
\end{align}
where $\gamma$ is the jammer's duty cycle and equals $1,\frac{T}{K},\frac{D}{K}$, or~$\alpha$
for barrage, pilot, data, or sparse jammers, respectively. This allows us to either precisely 
control the jammer energy (for jammers which are assumed to be essentially energy-limited)
or the transmit intensity (for jammers which may want to pass themselves off as a legitimate UE, for instance).
\vspace{-1mm}

\subsection{Performance Baselines} \label{sec:baseline}
We set the number of iterations for MAED and \mbox{SO-MAED} to $t_{\max}=20$ and 
emphasize again that we use only two different sets of weights for \mbox{SO-MAED}: one for QPSK transmission and one
for 16-QAM transmission. Neither \mbox{SO-MAED} nor MAED is adapted to the different jammer scenarios.
We compare our algorithms to the following baseline methods: 
The first baseline is the ``LMMSE'' method from \fref{sec:example}, which does not mitigate the jammer and 
separately performs least-squares (LS) channel estimation and LMMSE data detection.
The second baseline is the ``geniePOS'' method from \fref{sec:example}, which projects the receive
signals onto the orthogonal complement of the  true  jammer subspace 
and then separately performs LS channel estimation and LMMSE data detection in this projected space.
The last baseline, \mbox{``JL-SIMO,''} serves as a bound for attainable error-rate performance. 
This method operates in a jammerless but otherwise equivalent system and implements (with perfect channel knowledge) 
the single-input multiple-output (SIMO) bound
corresponding to the idealized case in which no inter-user interference is present.

\begin{figure*}[tp]
\!\!\!\!\!
\subfigure[strong barrage jammer (J1)]{
\includegraphics[height=4cm]{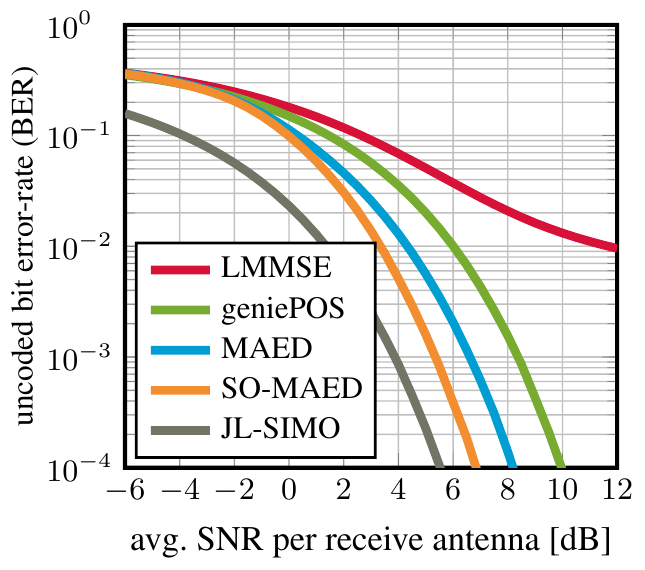}
\label{fig:qpsk:strong:static}
}\!\!
\subfigure[strong pilot jammer (J2)]{
\includegraphics[height=4cm]{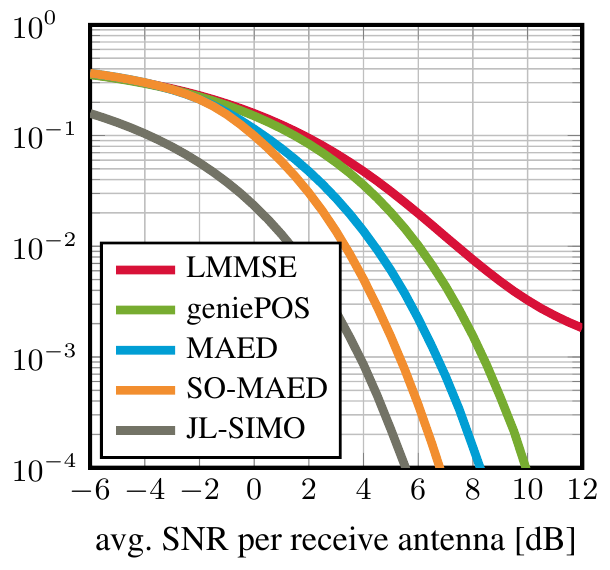}
\label{fig:qpsk:strong:pilot}
}\!\!
\subfigure[strong data jammer (J3)]{
\includegraphics[height=4cm]{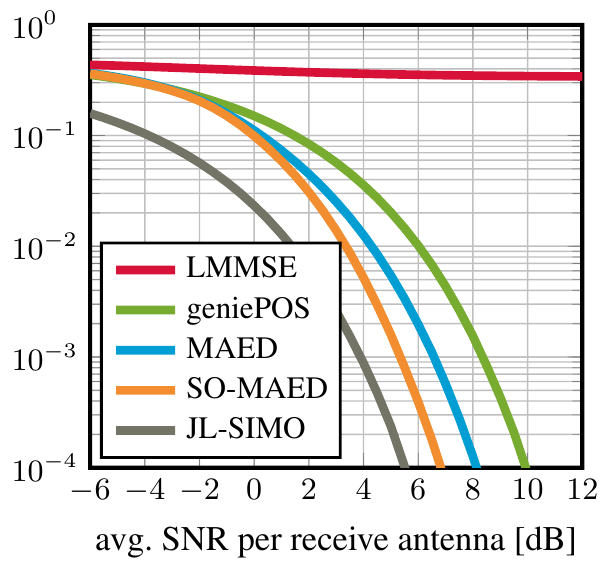}
\label{fig:qpsk:strong:data}
}\!\!
\subfigure[strong sparse jammer (J4)]{
\includegraphics[height=4cm]{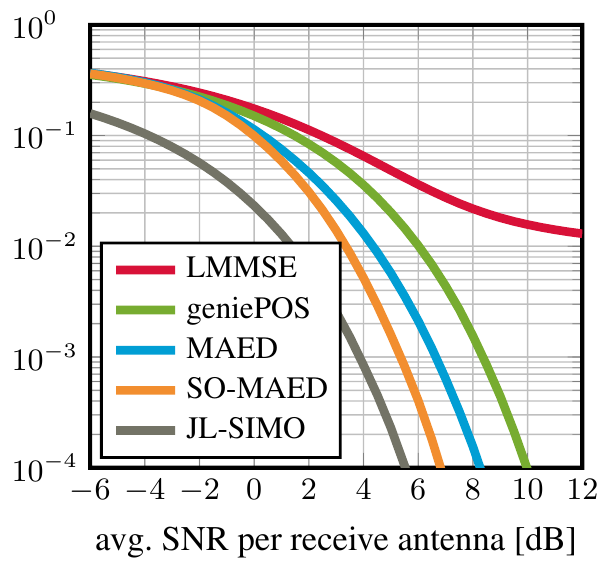}
\label{fig:qpsk:strong:burst}
}\!\!
\caption{Uncoded bit error-rate (BER) for \emph{QPSK} transmission in the presence of a \emph{strong} ($\rE=30$\,dB) jammer which
transmits Gaussian symbols (a) during the entire coherence interval,
(b) during the pilot phase only, (c) during the data phase only, or (d) in random unit-symbol bursts 
with a duty cycle of $\alpha=20\%$. 
}
\label{fig:qpsk_strong_jammers}
\end{figure*}
\begin{figure*}[tp]
\vspace{-1mm}
\!\!\!\!\!
\subfigure[strong barrage jammer (J1)]{
\includegraphics[height=4cm]{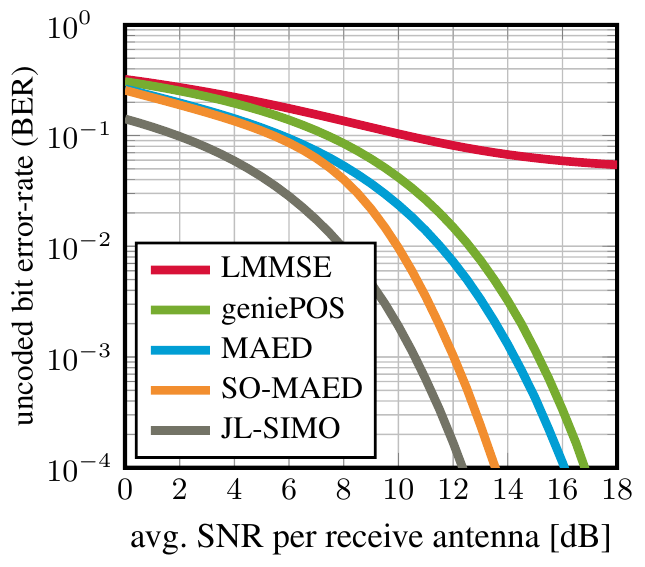}
\label{fig:strong:static}
}\!\!
\subfigure[strong pilot jammer (J2)]{
\includegraphics[height=4cm]{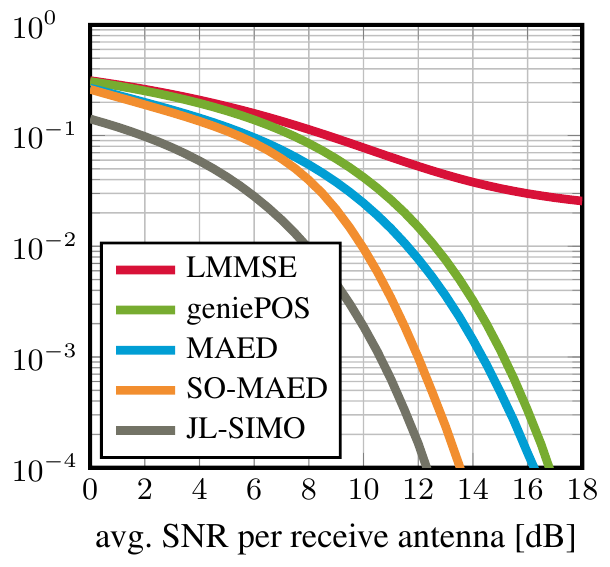}
\label{fig:strong:pilot}
}\!\!
\subfigure[strong data jammer (J3)]{
\includegraphics[height=4cm]{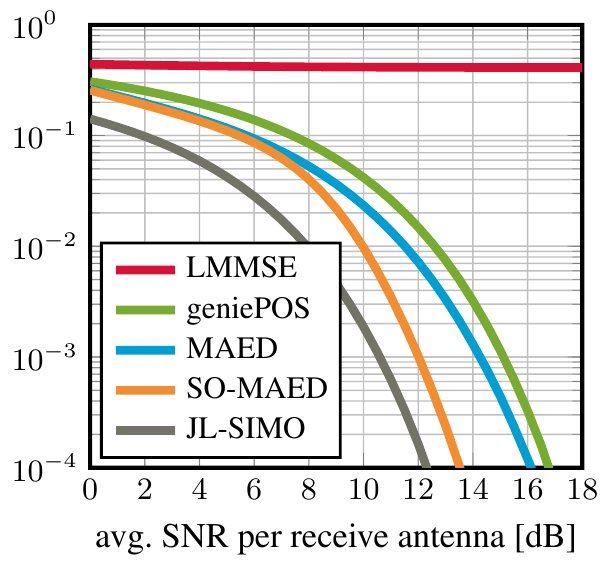}
\label{fig:strong:data}
}\!\!
\subfigure[strong sparse jammer (J4)]{
\includegraphics[height=4cm]{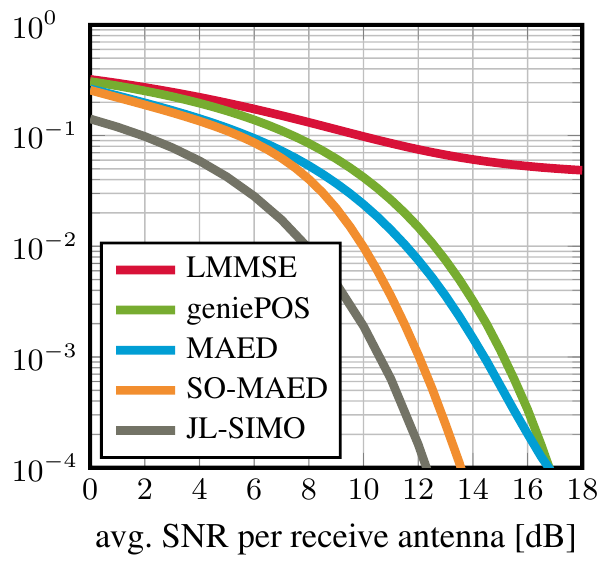}
\label{fig:strong:burst}
}\!\!
\caption{Uncoded bit error-rate (BER) for \emph{16-QAM} transmission in the presence of a \emph{strong} ($\rE\!=\!30$\,dB) jammer which transmits Gaussian symbols~(a)~during the entire coherence interval,
(b) during the pilot phase only, (c) during the data phase only, or (d) in random unit-symbol bursts 
with a duty cycle of $\alpha=20\%$. 
}
\label{fig:strong_jammers}
\end{figure*}
\begin{figure*}[h!]
\vspace{-1mm}
%
\!\!\!\!\!
\subfigure[weak barrage jammer (J1)]{
\includegraphics[height=4cm]{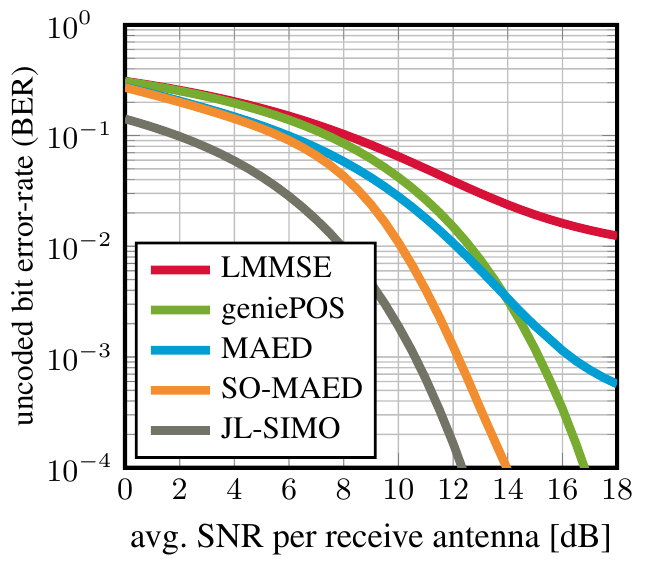}
\label{fig:strong:static}
}\!\!
\subfigure[weak pilot jammer (J2)]{
\includegraphics[height=4cm]{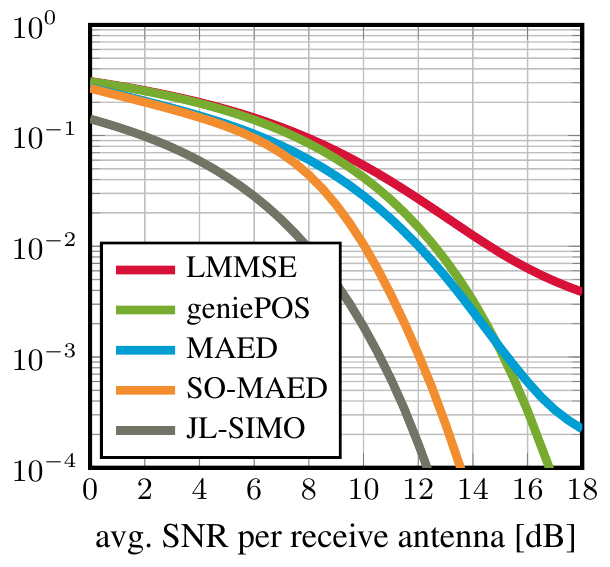}
\label{fig:strong:pilot}
}\!\!
\subfigure[weak data jammer (J3)]{
\includegraphics[height=4cm]{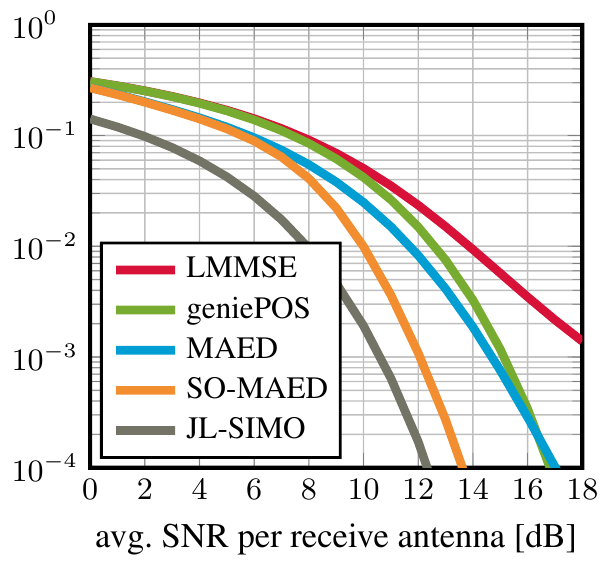}
\label{fig:strong:data}
}\!\!
\subfigure[weak sparse jammer (J4)]{
\includegraphics[height=4cm]{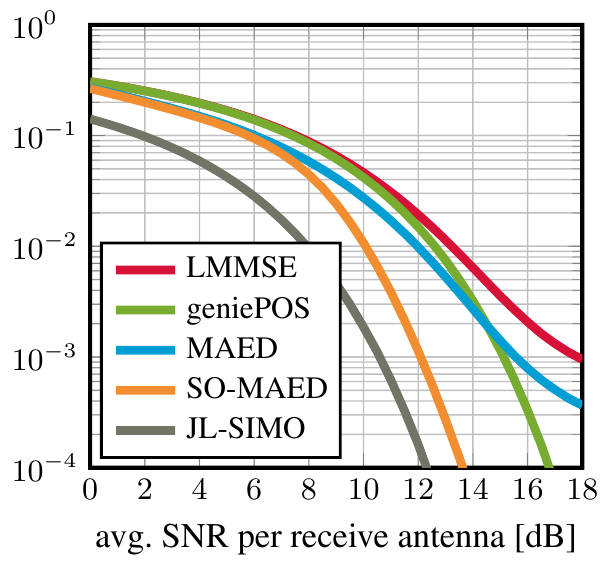}
\label{fig:strong:burst}
}\!\!
\caption{Uncoded bit error-rate (BER) for \emph{16-QAM} transmission in the presence of a \emph{weak} ($\rP=0$\,dB) jammer
which transmits 16-QAM symbols (a) during the entire coherence interval,
(b) during the pilot phase only, (c) during the data phase only, or (d) in random unit-symbol bursts 
with a duty cycle of $\alpha=20\%$. 
}
\label{fig:weak_jammers}
\end{figure*}

\begin{figure*}[tp]
\!\!\!\!\!
\subfigure[barrage jammers (J1)]{
\includegraphics[height=4cm]{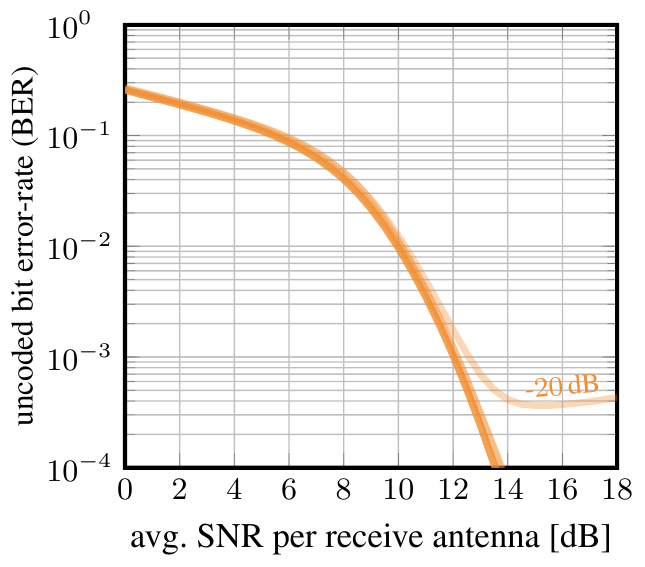}
\label{fig:many:static}
}\!\!
\subfigure[pilot jammers (J2)]{
\includegraphics[height=4cm]{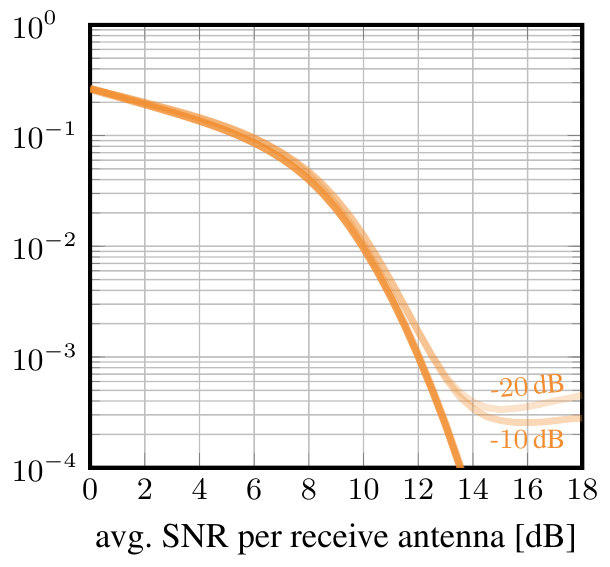}
\label{fig:many:pilot}
}\!\!
\subfigure[data jammers (J3)]{
\includegraphics[height=4cm]{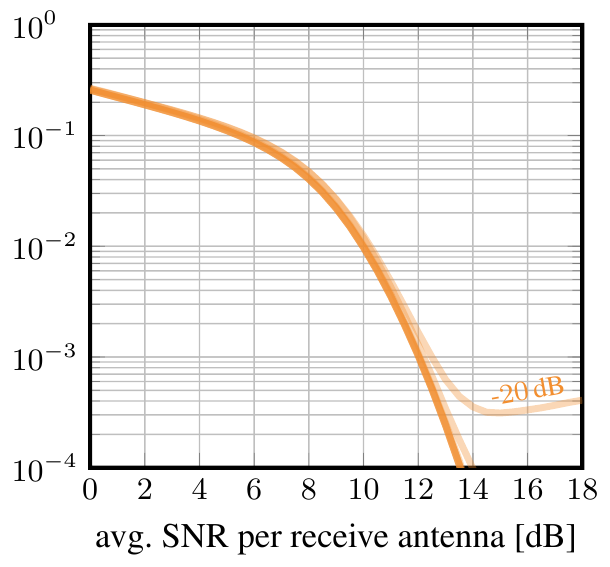}
\label{fig:many:data}
}\!\!
\subfigure[sparse jammers (J4)]{
\includegraphics[height=4cm]{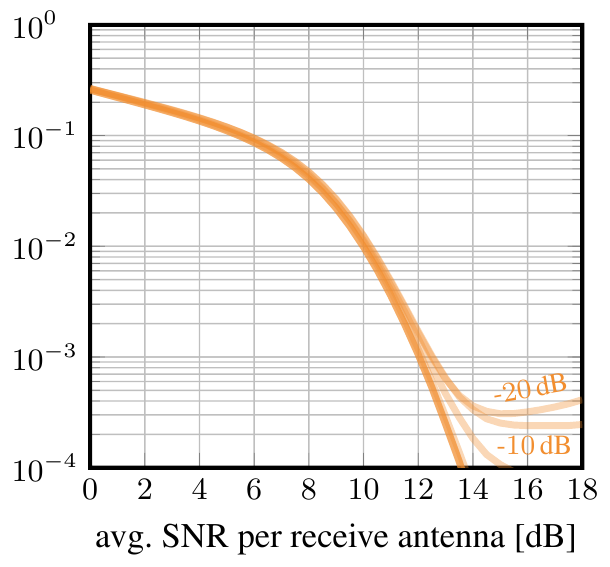}
\label{fig:many:burst}
}\!\!
\caption{Uncoded bit error-rate (BER) performance curves of SO-MAED in the presence of jammers with different receive powers
compared to the average UE,
$\rP\in\{-20\,\text{dB}, -10\,\text{dB}, 0\,\text{dB}, 10\,\text{dB}, 20\,\text{dB}, 40\,\text{dB}, 80\,\text{dB}\}$. 
The subfigures correspond to the different jammer types (J1)\,-\,(J4) and show one curve 
per~jammer power (plotted with 25\% opacity to depict the degree of overlap between curves). 
Curves that level off into an error floor are labeled with their jammer power, e.g.,
in \fref{fig:many:static}, the barrage jammer with receive power $\rP=-20$\,dB has an error floor while all other barrage jammers
have virtually identical BER curves.
}
\label{fig:many_jammers}
\end{figure*}

\subsection{Mitigation of Strong Gaussian Jammers}
We first investigate the ability of MAED and SO-MAED to mitigate strong jamming attacks. 
For this, we simulate Gaussian  jammers with \mbox{$\rE=30$\,dB} of all four types introduced in Section~\ref{sec:setup} 
and evaluate the performance of our algorithms compared to the baselines of Section~\ref{sec:baseline} for QPSK transmission (\fref{fig:qpsk_strong_jammers}) 
as well as for 16-QAM transmission (\fref{fig:strong_jammers}).
We note at this point that the performances of geniePOS and JL-SIMO are independent of the considered jammer type: geniePOS uses the genie-provided
jammer channel to null the jammer perfectly, regardless of its transmit sequence, and JL-SIMO operates on a jammerless system.
Unsurprisingly, the jammer-oblivious LMMSE baseline performs significantly worse than the jammer-robust geniePOS baseline under all attack scenarios,
with the data jamming attack turning out to be the most harmful and the pilot jamming attack the least harmful.
Both MAED and SO-MAED succeed in mitigating all four jamming attacks with highest effectiveness, even outperforming the genie-assisted geniePOS method by a
considerable margin.\footnote{The potential for MAED and SO-MAED to outperform geniePOS is a consequence of the superiority of joint channel estimation and data detection over 
separating channel estimation from data detection.}
Their efficacy is further reflected in the fact that SO-MAED and MAED approach the performance of the jammerless and MU interference-free
JL-SIMO bound to within less than $2$\,dB and $3$\,dB at $0.1$\%~BER, respectively, in all considered~scenarios.

The behavior is largely similar when 16-QAM instead of QPSK is used as transmit constellation~(\fref{fig:strong_jammers}). 
However, due to the decreased informativeness of the box prior for such higher-order constellations, MAED performs now closer 
to geniePOS, while SO-MAED still performs within $2$\,dB (at $0.1$\% BER) of the JL-SIMO bound.
The increased performance gap between them notwithstanding, both MAED and SO-MAED are able to  effectively mitigate all four attack types.

\subsection{Mitigation of Weak Constellation Jammers} \label{sec:results:weak}
We now turn to the analysis of more restrained jamming attacks in which the jammer transmits constellation symbols 
with relative power $\rP=0$\,dB during its on-phase (to pass itself off as just another UE, for instance \cite{vinogradova16a}).
Simulation results for 16-QAM transmission under all four types of jamming attacks are shown in~\fref{fig:weak_jammers}. 
Because of the weaker jamming attacks, the jammer-oblivious LMMSE baseline now performs closer to the jammer-resistant geniePOS baseline 
than it does in \fref{fig:strong_jammers}.
MAED again mitigates all attack types rather successfully, outperforming geniePOS in the low-SNR regime but slightly leveling off at high SNR.
Interestingly, MAED shows worse performance under these weak jamming attacks than under the strong jamming attacks of \fref{fig:strong_jammers}.
The reason is the following: MAED searches for the jamming subspace by looking for the dominant dimension 
of the iterative residual error $\tilde\bE^{(t)}$, see~\fref{eq:rayleigh}. If the received jamming energy is small compared to the 
received signal energy, then it becomes hard to distinguish the residual errors caused by the jamming signal from those caused by 
errors in estimating the channel and data matrices $\tilde\bH_\bP^{(t)}$ and $\tilde\bS_D^{(t)}$.
Note in contrast that, due to its superior signal prior, the equivalent performance loss of SO-MAED is only so small as to be 
virtually unnoticeable. Thus, SO-MAED outperforms MAED by a large margin and still approaches the JL-SIMO bound by less than 2dB at a BER of 0.1\%.\footnote{We 
\blue{note} that MAED does not suffer such a performance loss under weak jamming attacks when the transmit constellation is QPSK, 
since in that case the box signal prior of MAED is sufficiently accurate, \blue{cf.} \cite{marti2022smart}.}

\subsection{How Universal is SO-MAED Really?}
In the remainder of our evaluation, we focus mostly on \mbox{SO-MAED}, since it is clearly the better of the two proposed algorithms. 
To show that our approach indeed succeeds in mitigating arbitrary jamming attacks without need for fine-tuning of the algorithm or its parameters, 
\fref{fig:many_jammers} depicts performance results for a series of jamming attacks spanning a dynamic range from $\rP=-20$\,dB to 
$\rP=80$\,dB. Specifically, \fref{fig:many_jammers} shows results for all four jammer types, where every 
subfigure plots BER curves for jamming attacks with 
$\rP\in\{-20\,\text{dB}, -10\,\text{dB}, 0\,\text{dB}, 10\,\text{dB}, 20\,\text{dB}, 40\,\text{dB}, 80\,\text{dB}\}$.
The purpose of these plots is to illustrate that, apart from jamming attacks where the jammer is significantly weaker than 
the average UE\footnote{A jammer that is much weaker than the average UE resembles
a non-transmitting, and thus eclipsed, jammer; see Sections \ref{sec:theory} and \ref{sec:results:eclipsed}.}, 
the curves are virtually indistinguishable, meaning that the performance of SO-MAED is essentially independent of the specific
type of jamming attack that it is facing.

\subsection{Eclipsed Jammers} \label{sec:results:eclipsed}

\begin{figure}[tp]
\centering
\!\!\!\!\!
\subfigure[no jammer]{
\includegraphics[height=3.95cm]{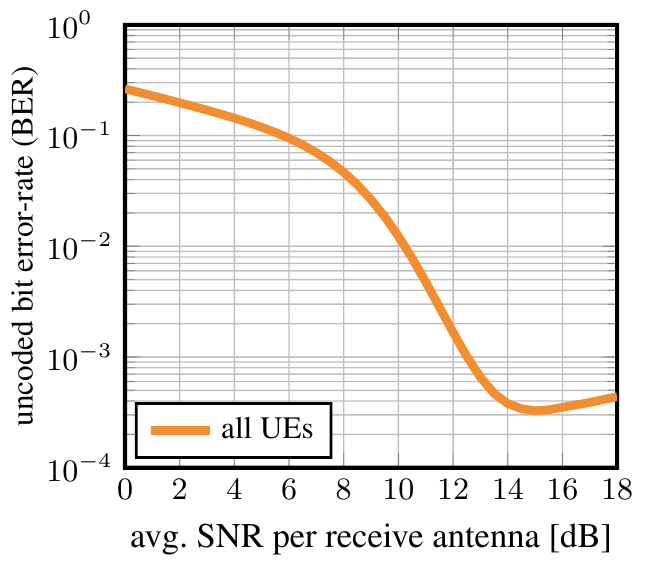}
\label{fig:eclipsed:no_jam}
}\!\!\!\!\!
\subfigure[weak UE-impersonating jammer]{
\includegraphics[height=3.95cm]{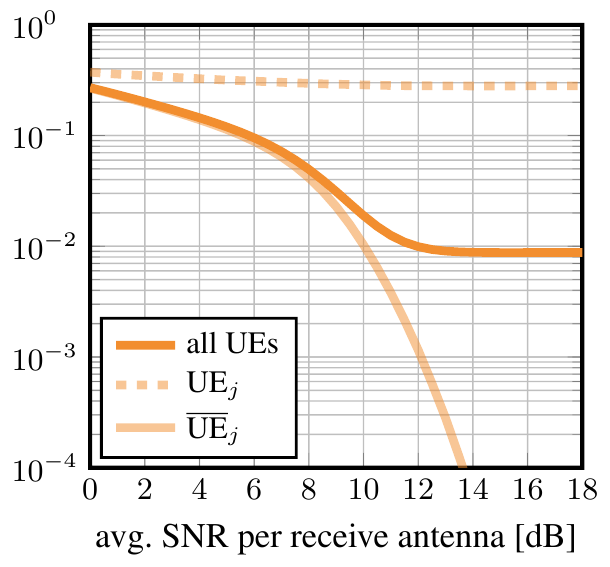}
\label{fig:eclipsed:weak_impers}
}\!\! \\
\!\!\!\!\!\!
\subfigure[strong UE-impersonating jammer]{
\includegraphics[height=3.95cm]{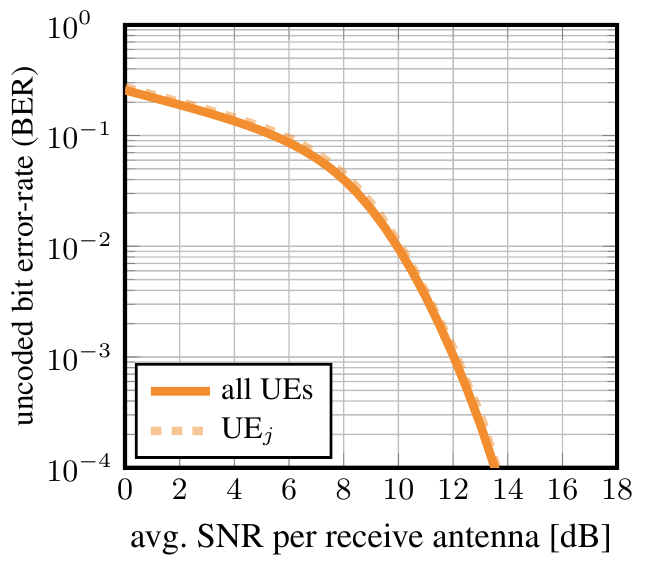}
\label{fig:eclipsed:strong_impers}
}\!\!\!
\subfigure[data-dependent jammer]{
\includegraphics[height=3.95cm]{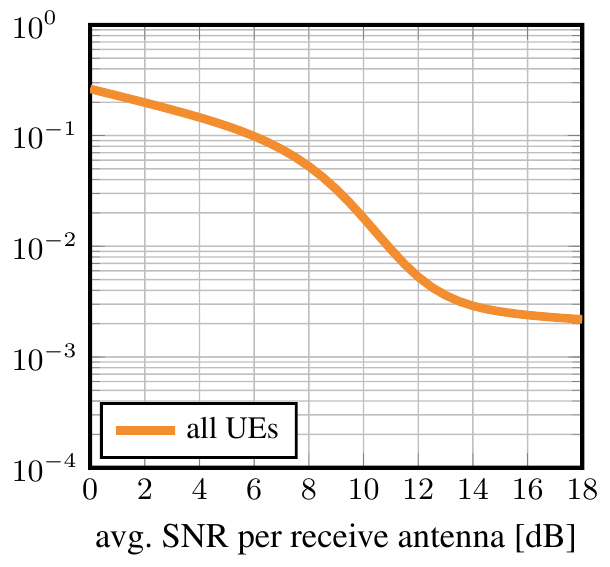}
\label{fig:eclipsed:row_eclipsed}
}\!\!
\caption{Uncoded bit error-rate of SO-MAED for different types of eclipsed jammers: (a) no jammer, 
(b) $\rP=0$\,dB jammer impersonating the $j$th UE by transmitting its pilot sequence 
($\text{UE}_j$ denotes the BER of the impersonated UE, and $\overline{\text{UE}}_j$ the BER among all other UEs),
(c) $\rP=30$\,dB jammer impersonating the $j$th UE by transmitting its pilot sequence, 
and (d) jammer causes eclipsing by transmitting a jamming sequence that depends on the UE transmit matrix~$\bS$.
Dashed lines represent the BER of the impersonated UE, transparent lines represent the BER among 
the UEs that are not impersonated by the jammer.
}
\label{fig:eclipsed}
\vspace{-2mm}
\end{figure}

Up to this point, the jamming signal $\bmw$ has always been \blue{independent of} the UE transmit matrix $\bS$.
The strong performance results of both MAED and SO-MAED have supported the claim in Remark~\ref{rem:rare} that,
in this case, eclipsing is the (rare) exception, not the norm.  
We now turn to an empirical analysis of how SO-MAED behaves when eclipsing \emph{does}
occur (\fref{fig:eclipsed}). To this end, we consider scenarios in which the jammer is eclipsed 
because there is no jamming activity (\fref{fig:eclipsed:no_jam}), 
because the jammer transmits a UE's pilot sequence (\fref{fig:eclipsed:weak_impers}, \fref{fig:eclipsed:strong_impers}),
or because the jamming sequence~$\bmw$ depends on the transmit matrix $\bS$ (which in reality would be unknown to the jammer) 
in a way that causes eclipsing (\fref{fig:eclipsed:row_eclipsed}).

In the case of no jammer (\fref{fig:eclipsed:no_jam}), or no jamming activity within a coherence interval,
we see that SO-MAED still reliably detects the transmit data. However, our method now suffers from an error floor
(albeit significantly below $0.1$\% BER). The reason for this error floor is that, in the absence of 
jamming energy to guide the choice of the nulled direction $\tilde\bmp$, there is the temptation to instead ``cover up''
detection errors (similar to the phenomenon discussed in \fref{sec:results:weak}).
However, the low level of the error floor shows that this potential pitfall does not cause a systematic breakdown 
of SO-MAED.
We emphasize also that SO-MAED does not simply null the strongest UE. Such (degenerate) behavior would only occur
if one UE were \emph{far} stronger than the others. With any reasonable power control scheme, UE nulling is not an 
issue.\footnote{This is exemplified by our experiments with i.i.d. Rayleigh fading \blue{channels,} which also exhibit minor imbalances in receive power between different UEs. \blue{See} also our results in \fref{sec:beyond}, 
where we use $\pm1.5$\,dB power control.}
In the case of a jammer that impersonates the $j$th UE by transmitting its pilot sequence in the training phase 
and constellation symbols in the data phase, with the same power as the average~UE \blue{($\rP=0$\,dB)}, 
SO-MAED indeed suffers a performance breakdown (\fref{fig:eclipsed:weak_impers}). However, 
closer inspection shows that this error floor is caused solely by errors in detecting the symbols of the 
impersonated UE. 
This is not surprising: The jammer is statistically indistinguishable from the $j$th UE,
so that is impossible to reliably separate the UE transmit symbols from the fake jammer transmit symbols. 
In this regard, we refer again to the information-theoretic discussion of \mbox{\cite[Sec. V]{lapidoth1998reliable}}.
Such impersonation attacks could be forestalled by using encrypted pilots~\cite{basciftci2015securing}.
If the jammer transmits the $j$th UE's pilot sequence and constellation symbols, but with much more
power ($\rP=30$\,dB), then the iterative detection procedure of \mbox{SO-MAED} will separate the jammer subspace from the 
$j$th UE's subspace (\fref{fig:eclipsed:strong_impers}), since, being so much stronger than any UE, 
the jammer subspace will dominate the residual matrix $\tilde\bE^{(t)}$ in~\fref{eq:rayleigh}.
Finally, \fref{fig:eclipsed:row_eclipsed} shows results for a case where the jammer knows $\bS$
and \blue{selects an $\tilde\bS_D$ which differs} from $\bS_D$ in a single row (with valid constellation symbols in the differing row), 
so that $\textit{rank}(\bS_D-\tilde\bS_D)=1$. It \blue{then} draws $\bmw_T\sim\setC\setN(\mathbf{0},\bI_D)$ and
sets $\tp{\bsj_D} = \tp{\bsj_T}\pinv{\bS_T}\tilde{\bS}_D$ to cause eclipsing \blue{(cf. \fref{def:eclipse})}. 
The jammer strength is $\rP=30$\,dB. 
The results show an error floor at roughly 0.2\% BER caused by the presence of an alternative spurious solution.
However, the results in Figs.~\ref{fig:qpsk_strong_jammers}\,--\,\ref{fig:many_jammers} show 
that when the jammer has to select~$\bmw$ without knowing $\bS_D$\blue{---which will be the case in most practical scenarios, as we argued in \fref{sec:limitations}---}, such accidental eclipsing is 
extremely rare.

\subsection{Beyond i.i.d. Rayleigh Fading} \label{sec:beyond}
\begin{figure}[tp]
\centering
\includegraphics[width=0.9\columnwidth]{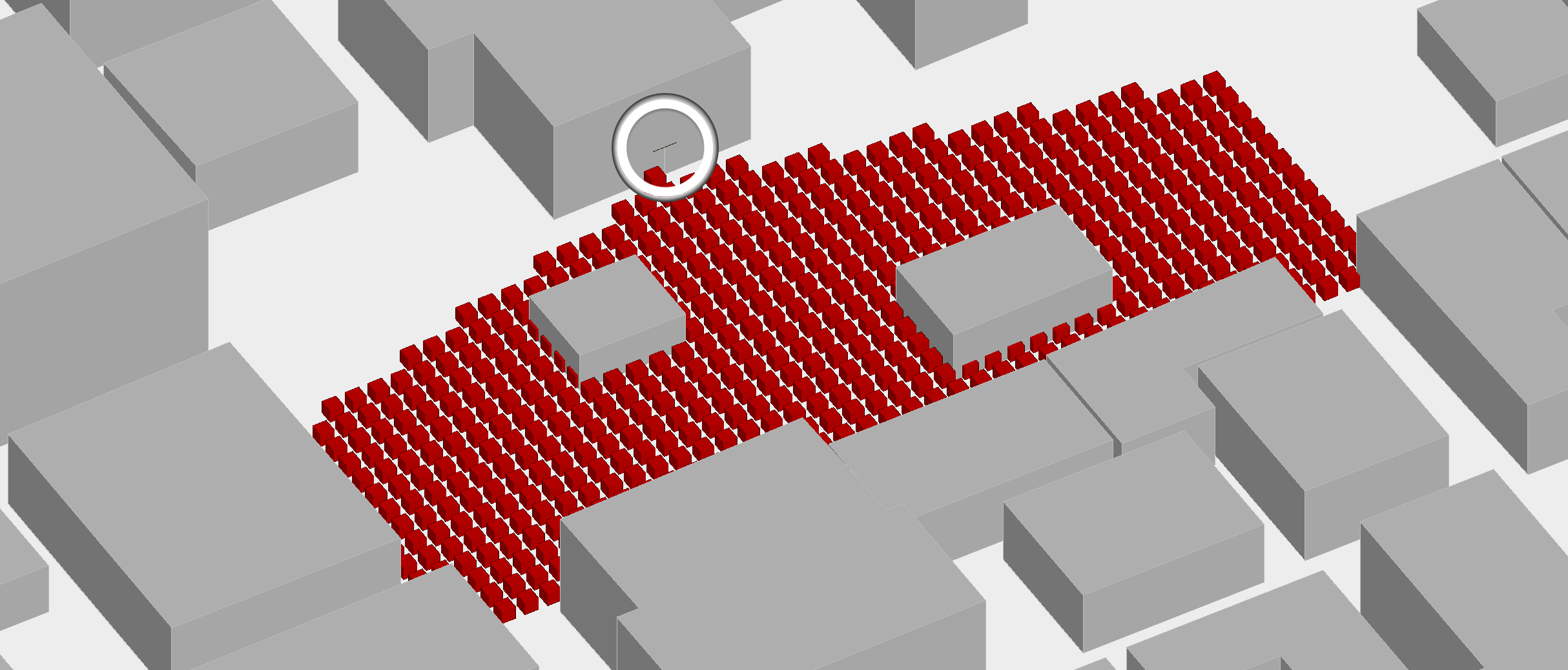}
\caption{Simulated scenario. The location of the BS his highlighted by the white circle while the red squares depict all possible UE locations.}
\label{fig:remcom_scenario}
\vspace{-2mm}
\end{figure}

So far, our experiments were based on i.i.d. Rayleigh fading channels, but our method does not depend
in any way on this particular channel model.  
To demonstrate that MAED and SO-MAED are also applicable in scenarios that deviate strongly from the i.i.d. Rayleigh model \blue{and that exhibit significant correlations between the
jammer's and the UEs' channels,}
we now evaluate our algorithms on mmWave channels generated 
with the commercial  Wireless InSite ray-tracer \cite{Remcom}. The simulated scenario is depicted in \fref{fig:remcom_scenario}.
We simulate a mmWave massive MU-MIMO system with a carrier frequency of $60$\,GHz and a bandwidth 
of $100$\,MHz. The BS is placed at a height of $10$\,m and consists of a horizontal uniform linear array with $B=128$ omnidirectional antennas spaced at half 
a wavelength. The omnidirectional single-antenna UEs and the jammer are located at a height of $1.65$\,m and placed in a $150^\circ$ sector 
spanning $180$\,m$\times$$90$\,m in front of the BS; see \fref{fig:remcom_scenario}. The~UEs~and the jammer are drawn at random from a grid with $5$\,m pitch while ensuring
that the minimum angular separation between any two UEs, as well as between the jammer and any UE, is $2.5^\circ$.
We assume $\pm1.5$\,dB power control, so that the ratio between the maximum and minimum per-UE receive power is~2.
\begin{figure}[tp]
\centering
\!\!\!\!\!
\subfigure[barrage jammer (J1)]{
\includegraphics[height=3.95cm]{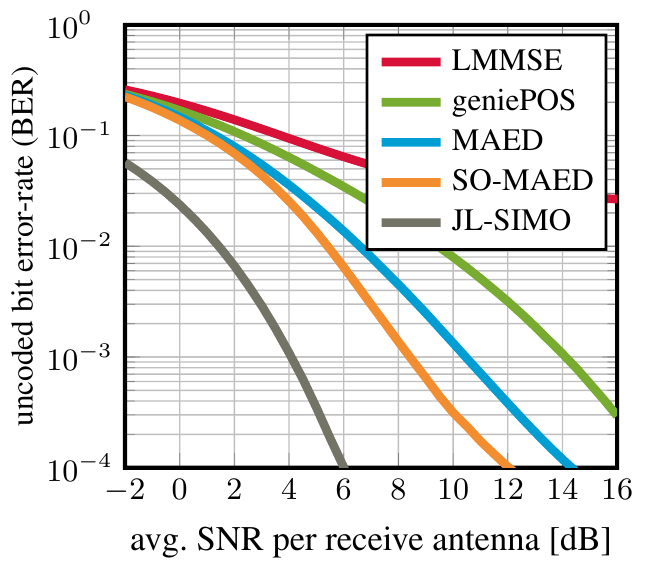}
\label{fig:remcom:barrage}
}\!\!\!
\subfigure[pilot jammer (J2)]{
\includegraphics[height=3.95cm]{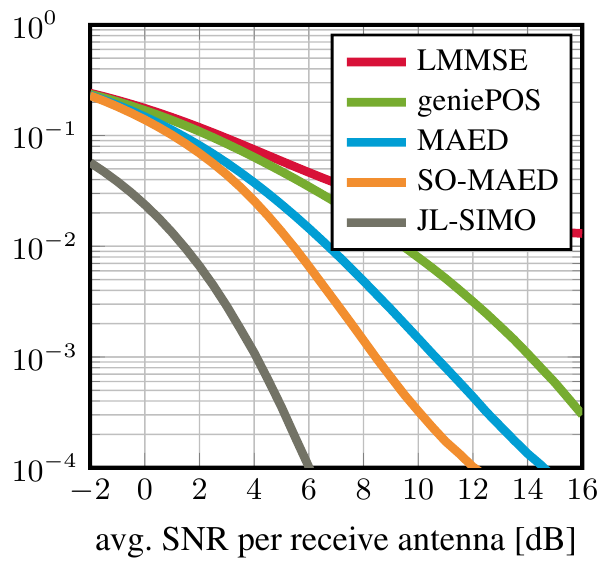}
\label{fig:remcom:pilot}
}\!\! \\
\!\!\!\!\!
\subfigure[data jammer (J3)]{
\includegraphics[height=3.95cm]{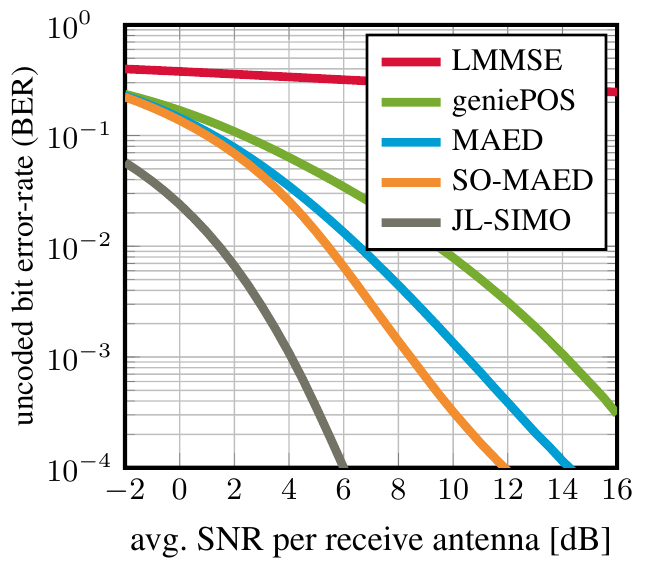}
\label{fig:remcom:data}
}\!\!\!
\subfigure[sparse jammer (J4)]{
\includegraphics[height=3.95cm]{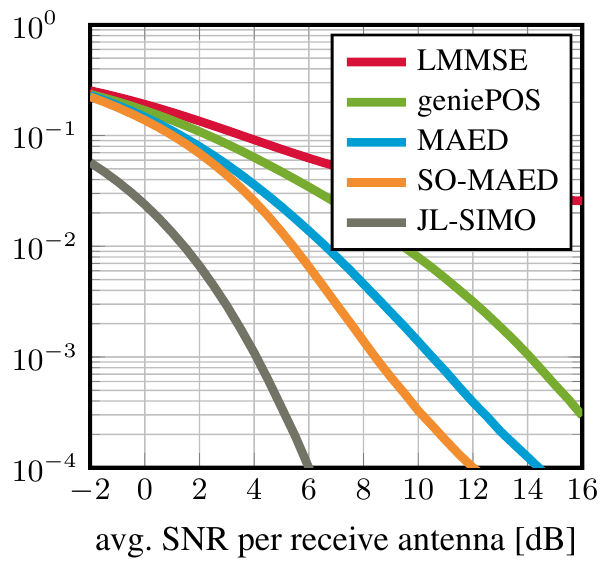}
\label{fig:remcom:sparse}
}\!\!
\caption{Uncoded bit error-rate (BER) for \emph{QPSK} transmission over realistic mmWave channels in the presence of a \emph{strong} ($\rE=30$\,dB) jammer.}
\label{fig:remcom_results}
\vspace{-2mm}
\end{figure}
The high correlation exhibited by these mmWave channels slows convergence of MAED and SO-MAED, so we increase their number of 
iterations to $t_{\max}=30$. We also retrain the parameters from SO-MAED on mmWave channels (while making a clear split between the training
set and the~evaluation~set).

The results for QPSK transmission in the presence of $\rE=30$\,dB are shown in \fref{fig:remcom_results}. 
The performance hierarchy is identical as in the equivalent Rayleigh-fading setup of \fref{fig:qpsk_strong_jammers}:
geniePOS is clearly outperformed by MAED, which is in turn outperformed by SO-MAED. 
However, the more challenging nature of mmWave channels amplifies performance differences: 
Due to its artificial immunity from the high inter-user interference of mmWave channels, JL-SIMO is now in a class of its own. 
However, MAED and SO-MAED gain almost $4$\,dB and $6$\,dB in SNR on geniePOS at $0.1\%$ BER, respectively, 
regardless of the jammer type. This shows that MAED and SO-MAED are also well suited 
for scenarios that deviate significantly from the i.i.d. Rayleigh model.
\vspace{-1mm}

\section{Conclusions}

We have proposed a method for the mitigation of smart jamming attacks on the massive
MU-MIMO uplink and supported its basic soundness with theoretical results. 
In contrast to existing mitigation methods, our approach does not rely on jamming activity 
during any particular time instant.
Instead, our method utilizes a newly proposed problem formulation which exploits the fact 
that the jammer's subspace remains constant within a coherence interval.
We have developed two efficient iterative algorithms, MAED and SO-MAED, which approximately solve the 
proposed optimization problem. 
Our simulation results have shown that MAED and SO-MAED are able to effectively mitigate a wide range of jamming attacks.
In particular, they succeed in mitigating attack types like data jamming and sparse jamming,
for which---to the best of our knowledge---no mitigation methods have existed so far.

\blue{There are numerous avenues for future work. 
Of particular importance is the development of methods for synchronization in the presence of smart jammers. 
Another issue is the question of how to obtain the required channel state information for transmit beamforming
in the case of multi-antenna UEs and point-to-point MIMO. 
Finally, our method focuses on the massive MIMO uplink. Equally important is, of course, the downlink, 
which presents the additional difficulty that UEs would probably only be able to rely on hybrid beamforming.
Since MAED and SO-MAED presume fully digital beamforming, the development of hybrid beamformers to mitigate 
smart jammers is therefore also a relevant problem for future work.}

\appendices

\vspace{-1mm}

\section{Proof of \fref{thm:maed}} \label{app:proof1}
\blue{Clearly,} if $\{\hat\bmp, \hat\bH_\bP, \hat\bS_D\}=\{\bmp, \bP\bH, \bS_D\}$, then
\begin{align}
	\hat\bP\bY - \hat\bH_\bP \hat\bS 
	&= \bP\bY - \bP\bH \bS \\
	&= \bP(\bH\bS + \Hj\tp{\bsj}) - \bP\bH \bS \\
	&= \bP \Hj\tp{\bsj} = \mathbf{0}, 
\end{align}
and so the objective \eqref{eq:obj1} is zero. Since the objective \mbox{function} is nonnegative, 
it follows that $\{\hat\bmp, \hat\bH_\bP, \hat\bS_D\}=\{\bmp, \bP\bH, \bS_D\}$ is a solution to \eqref{eq:obj1}. 
It remains to prove uniqueness. For this, we rewrite the objective in \eqref{eq:obj1} as
\begin{align}
	& \big\|\tilde\bP\bY - \tilde\bH_\bP \tilde\bS \big\|^2_F \nonumber\\
	&= \big\|\tilde\bP\bY_T - \tilde\bH_\bP \bS_T \big\|^2_F + \big\|\tilde\bP\bY_D - \tilde\bH_\bP \tilde\bS_D \big\|^2_F.
	\label{eq:decomposition}
\end{align}
The objective can only be zero if both terms on the right-hand-side (RHS) of \eqref{eq:decomposition} are zero.
The first term is zero \blue{if and only~if}
\begin{align}
	& \tilde\bP\bY_T - \tilde\bH_\bP \bS_T = \mathbf{0},
\end{align}
which implies
\begin{align}
	\tilde\bH_\bP = \tilde\bP\bY_T \pinv{\bS_T}, \label{eq:optimal_h}
\end{align}
since $\bS_T$ has full row rank.
Plugging this back into the second term on the RHS of \eqref{eq:decomposition} gives
\begin{align}
	& \tilde\bP\bY_D - \tilde\bH_\bP \tilde\bS_D \\
	&= \tilde\bP(\bY_D - \bY_T \pinv{\bS_T} \tilde\bS_D) \\
	&= \tilde\bP \left(\bH\bS_D + \Hj\tp{\bsj_D} - (\bH\bS_T + \Hj\tp{\bsj_T}) \pinv{\bS_T} \tilde\bS_D \right) \\
	&= \tilde\bP \left(\bH [\bS_D- \tilde\bS_D] + \Hj[\tp{\bsj_D}  - \tp{\bsj_T}\pinv{\bS_T} \tilde\bS_D] \right). \label{eq:term_mat}
\end{align}
The second term on the RHS of \eqref{eq:decomposition} (and, hence, the objective) is zero if and only if
the matrix in \eqref{eq:term_mat} is the zero matrix.
The projector $\tilde{\bP}$ can null a matrix of (at most) rank one. 
It follows that the objective function in \eqref{eq:decomposition} can be zero only if 
\begin{align}
	\bH [\bS_D- \tilde\bS_D] + \Hj[\tp{\bsj_D}  - \tp{\bsj_T}\pinv{\bS_T} \tilde\bS_D] \label{eq:eclipsing_equation}
\end{align}
is a matrix of (at most) rank one. Since $\bH$ has full column rank and\blue{, by assumption,}
$\Hj$ is not included in the \blue{column space} of $\bH$, 
this requires that 
\blue{the matrix $[\bS_D- \tilde\bS_D; \tp{\bsj_D}  - \tp{\bsj_T}\pinv{\bS_T} \tilde\bS_D]$
has rank one. By our assumption that the jammer is not eclipsed, this can only happen 
if $\tilde\bS_D = \bS_D$, so the estimated data matrix coincides with the true data matrix.}
In that case,~\eqref{eq:term_mat} is
\begin{align}
	\tilde\bP \Hj[\tp{\bsj_D}  - \tp{\bsj_T}\pinv{\bS_T} \tilde\bS_D], 
\end{align}
which (again by the assumption that the jammer is not \blue{eclipsed}) is zero if and only if $\tilde\bmp$ is collinear with $\bmj$, 
meaning that $\tilde\bmp = \alpha \bmp, |\alpha|=1$. This means that also the estimated jammer subspace coincides with the
true jammer subspace.
Finally, plugging this value of $\tilde\bmp$ back into \eqref{eq:optimal_h} yields
\begin{align}
	\tilde\bH_\bP &= \tilde\bP\bY_T \pinv{\bS_T}  \\
	&= \tilde\bP (\bH\bS_T + \Hj\tp{\bsj_T}) \pinv{\bS_T} \\
	&= \tilde\bP\bH\bS_T \pinv{\bS_T} = \bH_\bP, 
\end{align}
showing that also the estimated channel matrix coincides with the projection of the true channel matrix. 
We have thereby shown that $\big\|\tilde\bP\bY - \tilde\bH_\bP \tilde\bS \big\|^2_F$ is zero if and only if
$\tilde\bS_D=\bS_D$, $\tilde\bmp = \alpha\bmp, |\alpha|=1$, and $\tilde\bH_\bP = \bH_\bP$.
\hfill $\blacksquare$

\section{Proof of \fref{thm:maed2}} \label{app:proof2}

\blue{
$\bS_T$ is unitary, so $\pinv{\bS_T}=\herm{\bS_T}$. 
The jammer eclipses if there exists a matrix 
$\tilde\bS_D\in\setS^{U\times D}, \tilde\bS_D \neq \bS_D$ such that the matrix
\begin{align}
\boldsymbol{\Sigma} = 
\begin{bmatrix}
	\bS_D - \tilde\bS_D \\ \tp{\bsj_D} -\tp{\bsj_T}\herm{\bS_T}\tilde{\bS}_D \label{eq:the_matrix}
\end{bmatrix}
\end{align}
has rank one, meaning that 
\begin{align}
\begin{bmatrix}
	\bS_D - \tilde\bS_D \\ \tp{\bsj_D} -\tp{\bsj_T}\herm{\bS_T}\tilde{\bS}_D
\end{bmatrix} 
= 
\begin{bmatrix}
	\bma \\ \alpha
\end{bmatrix}  \tp{\bmb} \label{eq:the_event}
\end{align}
for some $\bma\in\opC^U, \bmb\in\opC^D, \alpha\in\opC$. 
Whether such an $\tilde\bS_D$ exists depends on the realization of the 
random matrices $\bS_D$ and $\bS_T$. 
\newline \indent
We now decompose the probability that an $\tilde\bS_D$ exists for which \eqref{eq:the_event} holds
(i.e., the probability that the jammer eclipses)
into the sum of the probability that such an $\tilde\bS_D$ exists which has rank one, plus the probability that
such an $\tilde\bS_D$ exists whose rank exceeds one. 
The proof proceeds by showing that the probability of the first of these two events is ``small,''
and that the probability of the second event is zero. 
\newline \indent
We start by bounding the probability that there exists a rank-one $\tilde\bS_D$ which satisfies \eqref{eq:the_event}.
Clearly, this probability is bounded by the probability that there exists a rank-one $\tilde\bS_D$ 
which satisfies $\bS_D - \tilde\bS_D = \bma\tp{\bmb}$ for some $\bma\in\opC^U, \bmb\in\opC^D$.
The entries of $\bS_D - \tilde\bS_D$ lie within the set 
\begin{align}
	\Delta\setS \triangleq \{ s-\tilde s \,:\, s, \tilde s \in \setS\}.
\end{align} 
Without loss of generality, any rank-one matrix $\bS_D - \tilde\bS_D$ can therefore be represented
by choosing the entries of the vectors $\bma$ and $\bmb$ from sets with cardinality 
$|\Delta\setS| \leq |\setS|^2$. For instance, 
one could pick the entries of $\bmb$ from $\Delta\setS$, and the entries of $\bma$ 
would have to come from at most $|\Delta\setS|$ different scaling factors. 
Analogously, $\tilde\bS_D$ was assumed to be of rank one, $\tilde\bS_D = \bmc\tp{\bmd}$, 
and since the entries of $\tilde\bS_D$ have to lie within $\setS$, we can 
without loss of generality restrict the entries of $\bmc, \bmd$ to lie within sets of cardinality $|\setS|$. 
We may thus write
\begin{align}
	\bS_D - \tilde\bS_D = \bS_D - \bmc\tp{\bmd} = \bma\tp{\bmb}.
\end{align}
Thus, to cause eclipsing, a rank-one $\bS_D$ would need to have the form 
\begin{align}
	\bS_D = \bma\tp{\bmb} + \bmc\tp{\bmd}. \label{eq:form1}
\end{align}
Since all $|\setS|^{UD}$ possible realizations of $\bS_D$ are equiprobable, 
the probability of \eqref{eq:form1} being satisfied
can be bounded by bounding the number of different matrices
that have the structure of \eqref{eq:form1}.
There are at most $|\setS|^{2U}$ different choices for~$\bma$, 
at most $|\setS|^{2D}$ different choices for $\bmb$,
at most $|\setS|^U$ different choices for $\bmc$, 
and at most $|\setS|^D$ different choices for~$\bmd$. 
In total, there are therefore at most $|\setS|^{3U+3D}$
different matrices that have the form \eqref{eq:form1}, 
and hence at most $|\setS|^{3U+3D}$
different realizations of $\bS_D$ for which there exists a rank-one $\tilde\bS_D$ that causes eclipsing.
Each of these realizations has probability $|\setS|^{-UD}$, 
so the probability of eclipsing with a rank-one $\tilde\bS_D$ can be bounded~by 
\begin{align}
	|\setS|^{-UD} |\setS|^{3U+3D} = |\setS|^{3U} |\setS|^{-(U-3)D}.
\end{align}
\indent
It remains to bound the probability that there exists an~$\tilde\bS_D$ whose rank exceeds one
and which satisfies \eqref{eq:the_event}.
For this, we consider the last row of \eqref{eq:the_matrix} and define $\bmx \triangleq \bS_T\bsj_T^\ast$.
Since~$\bS_T$ is Haar distributed,\footnote{The uniform distribution 
over unitary matrices is called Haar distribution.} 
$\bmx$ is distributed uniformly over the complex $U$-dimensional sphere of radius 
$\|\bmw_T\|_2$ \mbox{\cite[p.\,16]{meckes2019random}}, which, by assumption, is greater than zero.
Furthermore, $\bmx$ is independent of $\bmw_D$ and $\bS_D$. 
We may therefore write $\bmx = \frac{\|\bmw_T\|_2}{\|\bmz\|_2}\bmz$, 
where the entries of $\bmz$ are i.i.d. circularly-symmetric complex Gaussians
with unit variance, and where~$\bmz$ is independent of $\bmw_D$ and $\bS_D$.
So we rewrite the last row as 
\begin{align}
	\tp{\bsj_D} -\tp{\bsj_T}\herm{\bS_T}\tilde{\bS}_D
	= \tp{\bsj_D} - \frac{\|\bmw_T\|_2}{\|\bmz\|_2} \herm{\bmz}\tilde\bS_D.
\end{align}
Furthermore, for every $\bS_D\in\setS^{U\times D}$, we define the finite set
\begin{align}
	\setB(\bS_D) \triangleq 
	\big\{ \bmb \!\in\! \Delta\setS^D \,:\,& \exists \tilde\bS_D \!\in\! \setS^{U\times D}\!\setminus\!\{\bS_D\}
	\,\exists \bma\!\in\!\opC^U \nonumber \\ 
	& \text{such that } \bS_D - \tilde\bS_D  = \bma\tp{\bmb}	\big\}. 
\end{align}
The probability that there exists an $\tilde\bS_D$ whose rank exceeds one and which satisfies
\eqref{eq:the_event} can therefore be bounded by the probability that 
\begin{align}
	\tp{\bsj_D} - \frac{\|\bmw_T\|_2}{\|\bmz\|_2} \herm{\bmz}\tilde\bS_D \in \setB(\bS_D) \label{eq:B_set}
\end{align}
for some $\tilde\bS_D$ of rank greater than one. Using the union bound, we can in turn 
bound this probability by the sum (over all $\bmb\in\setB(\bS_D)$) of probabilities that
\begin{align}
	&\tp{\bsj_D} - \frac{\|\bmw_T\|_2}{\|\bmz\|_2} \herm{\bmz}\tilde\bS_D = \tp{\bmb}, 
\end{align}
which would imply that $\herm{\tilde\bS_D}\bmz^\ast$ is collinear with $\bmw_D - \bmb$.
So we can further bound the probability by the sum (over $\bmb$) of probabilities that
$\herm{\tilde\bS_D}\bmz^\ast$ is collinear with $\bmw_D -\bmb$ 
(note that $\herm{\tilde\bS_D}\bmz^\ast$ is independent of $\bmw_D -\bmb$).
Remember that the entries of $\bmz$, and hence of $\bmz^\ast$, are i.i.d. circularly-symmetric complex Gaussians
with unit variance. So $\herm{\tilde\bS_D}\bmz^\ast$ is a circularly-symmetric complex Gaussian vector
with covariance matrix $\herm{\tilde\bS_D}\tilde\bS_D$. 
And since $\herm{\tilde\bS_D}$ has at least two linearly independent rows (since
$\tilde\bS_D$ was assumed to be of rank greater than one), $\herm{\tilde\bS_D}\bmz^\ast$ has at least
two imperfectly correlated entries. Hence, for any fixed $\bmb$, the probability that
$\herm{\tilde\bS_D}\bmz^\ast$ is collinear with $\bmw_D -\bmb$ is zero. And since there are only finitely
many different vectors $\bmb$ to consider, the probability that \eqref{eq:B_set} holds is zero. 
In other words, the probability is zero that there exists a $\tilde\bS_D$ whose rank exceeds one
such that the jammer eclipses.
From this, the result follows. 
\hfill $\blacksquare$
}

\balance


\balance


\begin{thebibliography}{10}
\providecommand{\url}[1]{#1}
\csname url@samestyle\endcsname
\providecommand{\newblock}{\relax}
\providecommand{\bibinfo}[2]{#2}
\providecommand{\BIBentrySTDinterwordspacing}{\spaceskip=0pt\relax}
\providecommand{\BIBentryALTinterwordstretchfactor}{4}
\providecommand{\BIBentryALTinterwordspacing}{\spaceskip=\fontdimen2\font plus
\BIBentryALTinterwordstretchfactor\fontdimen3\font minus
  \fontdimen4\font\relax}
\providecommand{\BIBforeignlanguage}[2]{{%
\expandafter\ifx\csname l@#1\endcsname\relax
\typeout{** WARNING: IEEEtran.bst: No hyphenation pattern has been}%
\typeout{** loaded for the language `#1'. Using the pattern for}%
\typeout{** the default language instead.}%
\else
\language=\csname l@#1\endcsname
\fi
#2}}
\providecommand{\BIBdecl}{\relax}
\BIBdecl

\bibitem{marti2022smart}
G.~Marti and C.~Studer, ``Mitigating smart jammers in {MU-MIMO} via joint
  channel estimation and data detection,'' in \emph{Proc. IEEE Int. Conf.
  Commun. (ICC)}, May 2022, pp. 1336--1342.

\bibitem{economist2021satellite}
\BIBentryALTinterwordspacing
``Satellite-navigation systems such as {GPS} are at risk of jamming,''
  \emph{The Economist}. [Online]. Available:
  \url{https://www.economist.com/science-and-technology/2021/05/06/satellite-navigation-systems-such-as-gps-are-at-risk-of-jamming}
\BIBentrySTDinterwordspacing

\bibitem{topgun}
{J. Kosinski \textit{et al.}}, ``{Top Gun: Maverick},'' Paramount Pictures,
  2022.

\bibitem{popovski2014ultra}
P.~Popovski, ``Ultra-reliable communication in {5G} wireless systems,'' in
  \emph{Proc. Int. Conf. 5G Ubiquitous Connectivity}, Nov. 2014, pp. 146--151.

\bibitem{pirayesh2022jamming}
H.~Pirayesh and H.~Zeng, ``Jamming attacks and anti-jamming strategies in
  wireless networks: A comprehensive survey,'' \emph{{IEEE} Commun. Surveys
  Tuts.}, vol.~9, no.~2, pp. 767--809, 2022.

\bibitem{marti2021snips}
G.~Marti, O.~Casta\~neda, and C.~Studer, ``Jammer mitigation via beam-slicing
  for low-resolution {mmWave} massive {MU-MIMO},'' \emph{{IEEE} Open J.
  Circuits Syst.}, vol.~2, pp. 820--832, 2021.

\bibitem{yan2016jamming}
Q.~Yan, H.~Zeng, T.~Jiang, M.~Li, W.~Lou, and Y.~T. Hou, ``Jamming resilient
  communication using {MIMO} interference cancellation,'' \emph{{IEEE} Trans.
  Inf. Forensics Security}, vol.~11, no.~7, pp. 1486--1499, Jul. 2016.

\bibitem{shen14a}
W.~{Shen}, P.~{Ning}, X.~{He}, H.~{Dai}, and Y.~{Liu}, ``{MCR} decoding: A
  {MIMO} approach for defending against wireless jamming attacks,'' in
  \emph{Proc. IEEE Conf. Commun. Netw. Security (CNS)}, Oct. 2014, pp.
  133--138.

\bibitem{hoang2021suppression}
L.~M. Hoang, J.~A. Zhang, D.~N. Nguyen, X.~Huang, A.~Kekirigoda, and K.-P. Hui,
  ``Suppression of multiple spatially correlated jammers,'' \emph{{IEEE} Trans.
  Veh. Technol.}, vol.~70, no.~10, pp. 10\,489--10\,500, 2021.

\bibitem{zeng2017enabling}
H.~Zeng, C.~Cao, H.~Li, and Q.~Yan, ``Enabling jamming-resistant communications
  in wireless {MIMO} networks,'' in \emph{Proc. IEEE Conf. Commun. Netw.
  Security (CNS)}, Oct. 2017, pp. 1--9.

\bibitem{vinogradova16a}
J.~{Vinogradova}, E.~{Bj\"ornsson}, and E.~G. {Larsson}, ``Detection and
  mitigation of jamming attacks in massive {MIMO} systems using random matrix
  theory,'' in \emph{Proc. IEEE Int. Workshop Signal Process. Advances Wireless
  Commun. (SPAWC)}, Jul. 2016.

\bibitem{do18a}
T.~T. {Do}, E.~{Bj\"ornsson}, E.~G. {Larsson}, and S.~M. {Razavizadeh},
  ``Jamming-resistant receivers for the massive {MIMO} uplink,'' \emph{{IEEE}
  Trans. Inf. Forensics Security}, vol.~13, no.~1, pp. 210--223, Jan. 2018.

\bibitem{akhlaghpasand20a}
H.~{Akhlaghpasand}, E.~{Bj\"ornsson}, and S.~M. {Razavizadeh}, ``Jamming
  suppression in massive {MIMO} systems,'' \emph{{IEEE} Trans. Circuits Syst.
  {II}}, vol.~68, no.~1, pp. 182--186, Jan. 2020.

\bibitem{akhlaghpasand20b}
H.~{Akhlaghpasand}, E.~{Bj\"ornsson}, and S.~{Razavizadeh}, ``{Jamming-robust}
  uplink transmission for spatially correlated massive {MIMO} systems,''
  \emph{{IEEE} Trans. Commun.}, vol.~68, no.~6, pp. 3495--3504, Jun. 2020.

\bibitem{marti2021hybrid}
G.~Marti, O.~Casta\~neda, S.~Jacobsson, G.~Durisi, T.~Goldstein, and C.~Studer,
  ``Hybrid jammer mitigation for all-digital {mmWave} massive {MU-MIMO},'' in
  \emph{Proc. Asilomar Conf. Signals, Syst., Comput.}, Nov. 2021, pp. 93--99.

\bibitem{wan2022robust}
F.~Wan, J.~Xu, and Z.~Zhang, ``Robust beamforming based on covariance matrix
  reconstruction in fda-mimo radar to suppress deceptive jamming,''
  \emph{Sensors}, vol.~22, no.~4, p. 1479, 2022.

\bibitem{darsena2022anti}
D.~Darsena and F.~Verde, ``Anti-jamming beam alignment in millimeter-wave
  {MIMO} systems,'' \emph{{IEEE} Trans. Commun.}, 2022, early access.

\bibitem{miller2010subverting}
R.~Miller and W.~Trappe, ``Subverting {MIMO} wireless systems by jamming the
  channel estimation procedure,'' in \emph{Proc. ACM Conf. Wireless Netw.
  Security}, Mar. 2010, pp. 19--24.

\bibitem{miller2011vulnerabilities}
R.~Miller and W.~Trappe, ``On the vulnerabilities of {CSI} in {MIMO} wireless
  communication systems,'' \emph{{IEEE} Trans. Mobile Comput.}, vol.~11, no.~8,
  pp. 1386--1398, Aug. 2011.

\bibitem{clancy2011efficient}
T.~C. Clancy, ``Efficient {OFDM} denial: Pilot jamming and pilot nulling,'' in
  \emph{Proc. IEEE Int. Conf. Commun. (ICC)}, Jun. 2011, pp. 1--5.

\bibitem{sodagari2012efficient}
S.~Sodagari and T.~C. Clancy, ``Efficient jamming attacks on {MIMO} channels,''
  in \emph{IEEE Int. Conf. Commun. (ICC)}, Jun. 2012, pp. 852--856.

\bibitem{lichtman2013vulnerability}
M.~Lichtman, J.~H. Reed, T.~C. Clancy, and M.~Norton, ``Vulnerability of {LTE}
  to hostile interference,'' in \emph{Proc. IEEE Global Conf. Signal Inf.
  Process.}, Dec. 2013, pp. 285--288.

\bibitem{lichtman20185g}
M.~Lichtman, R.~Rao, V.~Marojevic, J.~Reed, and R.~P. Jover, ``{5G NR} jamming,
  spoofing, and sniffing: Threat assessment and mitigation,'' in \emph{Proc.
  IEEE Int. Conf. Commun. Workshop (ICCW)}, May 2018, pp. 1--6.

\bibitem{lichtman2016lte}
M.~Lichtman, R.~Jover, M.~Labib, R.~Rao, V.~Marojevic, and J.~H. Reed,
  ``{LTE/LTE--A} jamming, spoofing, and sniffing: threat assessment and
  mitigation,'' \emph{{IEEE} Commun. Mag.}, vol.~54, no.~4, pp. 54--61, Apr.
  2016.

\bibitem{girke2019towards}
F.~Girke, F.~Kurtz, N.~Dorsch, and C.~Wietfeld, ``Towards resilient {5G}:
  Lessons learned from experimental evaluations of {LTE} uplink jamming,'' in
  \emph{IEEE Int. Conf. Commun. Workshop (ICCW)}, May 2019, pp. 1--6.

\bibitem{lapan2012jamming}
M.~J. La~Pan, T.~C. Clancy, and R.~W. McGwier, ``Jamming attacks against {OFDM}
  timing synchronization and signal acquisition,'' in \emph{Proc. IEEE Mil.
  Commun. Conf. (MILCOM)}, Oct. 2012, pp. 1--7.

\bibitem{el2017lte}
A.~El-Keyi, O.~Ureten, H.~Yanikomeroglu, and T.~Yensen, ``{LTE} for public
  safety networks: Synchronization in the presence of jamming,'' \emph{IEEE
  Access}, vol.~5, pp. 20\,800--20\,813, Oct. 2017.

\bibitem{vikalo2006efficient}
H.~Vikalo, B.~Hassibi, and P.~Stoica, ``Efficient joint maximum-likelihood
  channel estimation and signal detection,'' \emph{{IEEE} Trans. Wireless
  Commun.}, vol.~5, no.~7, pp. 1838--1845, Jul. 2006.

\bibitem{xu2008exact}
W.~Xu, M.~Stojnic, and B.~Hassibi, ``On exact maximum-likelihood detection for
  non-coherent {MIMO} wireless systems: a branch-estimate-bound optimization
  framework,'' in \emph{Proc. IEEE Int. Symp. Inf. Theory (ISIT)}, Jul. 2008,
  pp. 2017--2021.

\bibitem{kofidis2017joint}
E.~Kofidis, C.~Chatzichristos, and A.~L. de~Almeida, ``Joint channel
  estimation/data detection in {MIMO-FBMC/OQAM} systems---a tensor-based
  approach,'' in \emph{Proc. Eur. Signal Process. Conf. (EUSIPCO)}, Aug. 2017,
  pp. 420--424.

\bibitem{castaneda2018vlsi}
O.~Casta\~neda, T.~Goldstein, and C.~Studer, ``{VLSI designs for joint channel
  estimation and data detection in large SIMO wireless systems},'' \emph{IEEE
  Transactions on Circuits and Systems I: Regular Papers}, vol.~65, no.~3, pp.
  1120--1132, Mar. 2018.

\bibitem{yilmaz2019channel}
B.~B. Yilmaz and A.~T. Erdogan, ``Channel estimation for massive {MIMO}: A
  semiblind algorithm exploiting {QAM} structure,'' in \emph{Proc. Asilomar
  Conf. Signals, Syst., Comput.}, Nov. 2019, pp. 2077--2081.

\bibitem{he2020model}
H.~He, C.-K. Wen, S.~Jin, and G.~Y. Li, ``Model-driven deep learning for {MIMO}
  detection,'' \emph{{IEEE} Trans. Signal Process.}, vol.~68, pp. 1702--1715,
  Feb. 2020.

\bibitem{song2021soft}
H.~Song, X.~You, C.~Zhang, and C.~Studer, ``Soft-output joint channel
  estimation and data detection using deep unfolding,'' in \emph{Proc. IEEE
  Inf. Theory Workshop (ITW)}, Oct. 2021, pp. 1--5.

\bibitem{goldstein16a}
\BIBentryALTinterwordspacing
T.~Goldstein, C.~Studer, and R.~G. Baraniuk, ``A field guide to
  forward-backward splitting with a {FASTA} implementation,'' Feb. 2016.
  [Online]. Available: \url{https://arxiv.org/abs/1411.3406}
\BIBentrySTDinterwordspacing

\bibitem{hershey2014deep}
J.~R. Hershey, J.~L. Roux, and F.~Weninger, ``Deep unfolding: Model-based
  inspiration of novel deep architectures,'' \emph{arXiv:1409.2574}, 2014.

\bibitem{balatsoukas2019deep}
A.~Balatsoukas-Stimming and C.~Studer, ``Deep unfolding for communications
  systems: A survey and some new directions,'' in \emph{Proc. IEEE Int.
  Workshop Signal Process. Syst. (SiPS)}.\hskip 1em plus 0.5em minus
  0.4em\relax IEEE, 2019, pp. 266--271.

\bibitem{goutay2020deep}
M.~Goutay, F.~A. Aoudia, and J.~Hoydis, ``Deep hypernetwork-based {MIMO}
  detection,'' in \emph{Proc. IEEE Int. Workshop Signal Process. Advances
  Wireless Commun. (SPAWC)}, 2020, pp. 1--5.

\bibitem{monga2021algorithm}
V.~Monga, Y.~Li, and Y.~C. Eldar, ``Algorithm unrolling: Interpretable,
  efficient deep learning for signal and image processing,'' \emph{{IEEE}
  Signal Process. Mag.}, vol.~38, no.~2, pp. 18--44, 2021.

\bibitem{marti2022joint}
G.~Marti and C.~Studer, ``Joint jammer mitigation and data detection for smart,
  distributed, and multi-antenna jammers,'' \emph{to be presented at IEEE Int. Conf. Commun. (ICC)}, 2023.

\bibitem{sabharwal2014band}
A.~Sabharwal, P.~Schniter, D.~Guo, D.~W. Bliss, S.~Rangarajan, and R.~Wichman,
  ``In-band full-duplex wireless: Challenges and opportunities,'' \emph{{IEEE}
  J. Sel. Areas Commun.}, vol.~32, no.~9, pp. 1637--1652, Sep. 2014.

\bibitem{GV96}
G.~H. Golub and C.~F. {van Loan}, \emph{Matrix Computations}, 3rd~ed.\hskip 1em
  plus 0.5em minus 0.4em\relax The Johns Hopkins Univ. Press, 1996.

\bibitem{lapidoth1998reliable}
A.~Lapidoth and P.~Narayan, ``Reliable communication under
  channel~un-certainty,'' \emph{{IEEE} Trans. Inf. Theory}, vol.~44, no.~6, pp.
  2148--2177, 1998.

\bibitem{jeon2021mismatched}
C.~Jeon, A.~Maleki, and C.~Studer, ``Mismatched data detection in massive
  {MU-MIMO},'' \emph{{IEEE} Trans. Signal Process.}, vol.~69, pp. 6071--6082,
  2021.

\bibitem{parikh13a}
N.~Parikh and S.~Boyd, ``Proximal algorithms,'' \emph{Found. Trends Optim.},
  vol.~1, no.~3, pp. 127--239, Jan. 2014.

\bibitem{barzilai1988two}
J.~Barzilai and J.~M. Borwein, ``Two-point step size gradient methods,''
  \emph{IMA J. Numer. Anal.}, vol.~8, no.~1, pp. 141--148, 1988.

\bibitem{collings2004low}
I.~B. Collings, M.~R. Butler, and M.~McKay, ``Low complexity receiver design
  for {MIMO} bit-interleaved coded modulation,'' in \emph{IEEE Int. Symp.
  Spread Spectrum Techniques Applicat.}, Aug. 2004, pp. 12--16.

\bibitem{Baydin2018automatic}
A.~G. Baydin, B.~A. Pearlmutter, A.~A. Radul, and J.~M. Siskind, ``Automatic
  differentiation in machine learning: a survey,'' \emph{Journal of Machine
  Learning Research}, vol.~18, no. 153, pp. 1--43, 2018.

\bibitem{tensorflow2015whitepaper}
\BIBentryALTinterwordspacing
{M. Abadi \textit{et al.}}, ``{TensorFlow}: Large-scale machine learning on
  heterogeneous systems,'' 2015. [Online]. Available:
  \url{https://www.tensorflow.org/}
\BIBentrySTDinterwordspacing

\bibitem{bussgang52a}
J.~J. Bussgang, ``Crosscorrelation functions of amplitude-distorted {Gaussian}
  signals,'' Cambridge, MA, Tech. Rep. 216, Mar. 1952.

\bibitem{minkoff85a}
J.~Minkoff, ``The role of {AM}-to-{PM} conversion in memoryless nonlinear
  systems,'' \emph{{IEEE} Trans. Commun.}, vol.~33, no.~2, pp. 139--144, Feb.
  1985.

\bibitem{tomasoni2006low}
A.~Tomasoni, M.~Ferrari, D.~Gatti, F.~Osnato, and S.~Bellini, ``A low
  complexity turbo {MMSE} receiver for {W-LAN MIMO} systems,'' in \emph{Proc.
  IEEE Int. Conf. Commun. (ICC)}, vol.~9, Jun. 2006, pp. 4119--4124.

\bibitem{kingma2014adam}
D.~P. Kingma and J.~Ba, ``Adam: A method for stochastic optimization,''
  \emph{arXiv preprint arXiv:1412.6980}, 2014.

\bibitem{keras_adam}
``Keras adam class,'' \url{https://keras.io/api/optimizers/adam/}, accessed:
  2022.

\bibitem{jagannath2021redefining}
A.~Jagannath, J.~Jagannath, and T.~Melodia, ``Redefining wireless communication
  for {6G}: Signal processing meets deep learning with deep unfolding,''
  \emph{IEEE Trans. Artificial Intelligence}, vol.~2, no.~6, pp. 528--536, Dec.
  2021.

\bibitem{albreem2021deep}
M.~A. Albreem, A.~H. Alhabbash, S.~Shahabuddin, and M.~Juntti, ``Deep learning
  for massive mimo uplink detectors,'' \emph{{IEEE} Commun. Surveys Tuts.},
  vol.~24, no.~1, pp. 741--766, 2021.

\bibitem{basciftci2015securing}
Y.~O. Basciftci, C.~E. Koksal, and A.~Ashikhmin, ``Securing massive {MIMO} at
  the physical layer,'' in \emph{Proc. IEEE Conf. Commun. Netw. Security
  (CNS)}, Sep. 2015, pp. 272--280.

\bibitem{Remcom}
``Remcom,'' \url{https://remcom.com/wireless-insite-em-propagation-software/},
  accessed: 2022.

\bibitem{meckes2019random}
E.~S. Meckes, \emph{The random matrix theory of the classical compact
  groups}.\hskip 1em plus 0.5em minus 0.4em\relax Cambridge University Press,
  2019, vol. 218.

\end{thebibliography}
\end{document}